\documentclass[a4paper,english,aps,pra,twocolumn,showpacs,amsmath]{revtex4}
\usepackage[T1]{fontenc}
\usepackage[latin1]{inputenc}
\usepackage{prettyref}
\usepackage{color}
\usepackage{graphicx}
\usepackage{amssymb}

\makeatletter

\usepackage{comment}
\renewcommand{\vec}[1]{\mathbf{#1}}

\usepackage{babel}
\makeatother
\begin{document}

\title{BCS-BEC Crossover in Atomic Fermi Gases with a Narrow Resonance}

\author{L. M. Jensen$^{1,2,3}$}

\author{H. M. Nilsen$^{4}$}

\author{Gentaro Watanabe$^{1,5}$}

\affiliation{$^{1}$NORDITA, Blegdamsvej 17, DK-2100, Copenhagen \O, Denmark\\
$^{2}$Department of Physics, Nanoscience Center, P.O. Box 35, FIN-40014 University of Jyv\"askyl\"a, Finland\\
$^{3}$Department of Theoretical Physics, Ume\aa\ University, Ume\aa, 901
87, Sweden\\
$^{4}$Center for Mathematics for Applications, P.O. Box 1053, Blindern,
NO-0316, Oslo Norway\\
$^{5}$The Institute of Chemical and Physical Research (RIKEN), 2-1 Hirosawa,
Wako, Saitama 351-0198, Japan}

\begin{abstract}
We determine the effects on the BCS-BEC crossover of the energy
dependence of the effective two-body interaction, which at low
energies is determined by the effective range. To describe
interactions with an effective range of either sign, we consider a
single-channel model  with a two-body interaction having an
attractive square well and a repulsive square barrier.  We
investigate the two-body scattering properties of the model, and 
then solve the Eagles-Leggett equations for the zero temperature 
crossover, determining the momentum  dependent gap and the chemical 
potential self-consistently. From this we investigate the dependence 
of the crossover on the effective range of the interaction.
\end{abstract}


\maketitle

\section{introduction\label{sec:introduction}}

Recently there has been remarkable progress in realizing, in
ultracold atomic Fermi gases, the crossover from weak-coupling
(BCS) superfluidity to Bose-Einstein condensation (BEC) of bound
diatomic molecules. The key to this development is that, due to
Feshbach molecular resonances, the strength and sign of the
effective interatomic interaction can be tuned by varying the
external magnetic field. For positive scattering lengths the
resonantly enhanced interaction has been used to create a
long-lived gas of diatomic molecules and Bose-Einstein condensates
of molecules. For a negative scattering length, experimental
evidence has been obtained for an atom-pair correlated state
analogous to the BCS superconducting state at weak coupling 
\cite{Regal2004,Zwierlein2004a}.
Additional studies of collective modes \cite{Kinast2004,Bartenstein2004col}
and the spectroscopic pairing gap \cite{Chin2004} 
in the crossover region also display behavior consistent with
pairing, the most convincing experiment to date being the
observation of a vortex lattice in the entire BCS-BEC crossover
region \cite{Zwierlein2005b}.

The simplest model describing the crossover is due to
Eagles and Leggett \cite{Eagles1969b,Leggett1980,Leggett1980b}, in
which the effective fermion-fermion interaction is parametrized by the
scattering length, $a$. This approximation for the interaction is
valid as long as the effective fermion-fermion interaction does
not vary significantly on the scale of the Fermi energy, $E_{\rm
F}$, the energy relevant for the many-body problem  \cite{Bruun2004a,
Petrov2004b}. In practice, large scattering lengths are realized
by using Feshbach resonances, and resonances for which this
condition holds are referred to as ``broad''.  For such
resonances, the atom-molecule coupling and the width parameter for
the resonance are large. In the opposite case, the resonance is
referred to as ``narrow''. Most experiments on the crossover to
date have been performed with broad resonances. For the resonance
in $^{6}\mathrm{Li}$, the experimental data on the evolution of
condensate profile \cite{Bartenstein2004a} agrees well with
the universal single-channel model \cite{Perali2004a,Simonucci2005}.
In \cite{Kinast2004,Bartenstein2004col} measurements of the collective axial
and radial modes of a trapped $^{6}\mathrm{Li}$ gas were reported
to be quantitatively in agreement with results obtained from the
zero temperature BCS-BEC crossover \cite{Heiselberg2004,Hu2004}.

In the future, one may anticipate that experiments will be made on the
crossover for narrow resonances. It is therefore of interest to
investigate the influence of energy dependence of the interaction
on the crossover. The leading contributions to the effective
interaction beyond the scattering length are expressed in terms of
the effective range, $r_e$, which is defined in terms of the
s-wave scattering phase shift $\delta$ by the equation
\begin{equation}
k\cot\delta=-\frac{1}{a}+\frac{1}{2}r_{e}k^{2}+O(k^{4}),
\label{eq:expansion}
\end{equation}
where $k$ is the wave number for the relative motion. The influence of the 
effective range is also interesting from the point of view
of the BCS-BEC crossover produced by varying the
density of the gas \cite{Andrenacci1999}. In this case, the effective
interaction between atoms at the Fermi surface depends on the Fermi
momentum, and hence depends on density. Several approaches to study 
effective range dependence are based on microscopic interaction 
potentials, e.g., coupled square
wells \cite{Kokkelmans2002}, the modified P\"oschl-Teller
potential \cite{Carlson2003}, and the Gaussian potential
\cite{Andrenacci1999,Parish2005a}. For monotonic attractive
potentials the effective range is positive and can be
shown to be related to an average range of the microscopic potential.
The effective range expansion applied to the
multichannel scattering problem yields effective ranges of either sign.
In Refs.\ \cite{Palo2004,Diener2004b} the properties of a square well plus
square barrier model potential were matched with the scattering
properties of a renormalized two-channel resonance 
model \cite{Holland2001} and the zero temperature atom-molecule
BCS-BEC crossover was re-expressed in terms of an effective
single-channel potential model with a large and negative effective
range \cite{Bruun2004a,Palo2004}.

The purpose of this paper is to investigate the dependence of the
crossover on the energy dependence of the effective two-body
interaction.  We shall start from a microscopic model, a
square-well, square-barrier potential, which has previously been
employed in Ref.\ \cite{Palo2004}. 
This can be used to construct effective interactions with different 
scattering lengths and an effective range that can be either positive 
or negative. We then solve the Eagles-Leggett equations,
determining the gap and the chemical potential self-consistently.
Non-trivial corrections due to the negative effective range have not 
been fully investigated so far and 
such calculations provide a useful guide to explore the BCS-BEC crossover 
for narrow resonances in the future.

The paper is organized as follows: Section
\ref{sec:Effective-range} contains a general introduction to the
low energy effective range expansion of the effective interatomic
interaction in a dilute Fermi gas \cite{Blatt1949,Bethe1949}. In
Section \ref{sec:sqwb} the two-body properties of the square-well,
square-barrier potential are derived.  Subsequently, in Section
\ref{sec:crossover} we numerically solve the zero temperature
BCS-BEC crossover problem using the Eagles-Leggett
variational state  \cite{Eagles1969b,Leggett1980} and re-express
the crossover properties in terms of the low energy two-body
properties.

\section{Effective range expansion\label{sec:Effective-range}}

Historically, the effective range expansion originated in attempts
to understand nuclear forces from low energy nucleon scattering experiments
\cite{Blatt1949,Bethe1949}.  Classical problems such
as nucleon-nucleon scattering, the deuteron spectrum, and $\alpha$
decay were studied by such methods. Recently, the effective range was
introduced to the field of ultracold atom gases where its size and sign
are related to the width of the Feshbach resonances.

The dominant scattering process in an ultracold two-component Fermi
gas is the binary elastic s-wave scattering between the atoms.
When the typical wavelength $\lambda$ corresponding to the collision
energy $E\sim\hbar^{2}/m\lambda^{2}$ ($m$ is the mass of the fermion
atom) is much larger than the range of the inter-atomic potential
$U(\vec{r})$, this process, in the center-of-mass frame, can be described
by the Schr\"{o}dinger equation with a radial wave function $r\psi$
of the s-state: 
\begin{equation}
-\frac{\hbar^{2}}{m}\vec{\nabla}^{2}\psi(\vec{r})+U(\vec{r})\psi(\vec{r})=E\psi(\vec{r}).\label{eq:schrodinger}
\end{equation}

Let us consider an interatomic potential with a finite range $r_{1}$,
i.e., $U(r)=0$ at $r>r_{1}$. Using the solutions $\psi_{1}$ and
$\psi_{2}$ of Eq. \prettyref{eq:schrodinger} with the energies $E_{i}=\hbar^{2}k_{i}^{2}/m$
($i=1,2$), we get 
\begin{equation}
\left[\psi_{2}\frac{d\psi_{1}}{dr}-\psi_{1}\frac{d\psi_{2}}{dr}\right]_{0}^{r_{1}}=(k_{2}^{2}-k_{1}^{2})\int_{0}^{r_{1}}\psi_{1}\psi_{2}dr.
\label{eq:semi_ortho}
\end{equation}
 Following Bethe \cite{Bethe1949}, we introduce a reference wave
function $\psi_{\mathrm{ref},i}$ which coincides with $\psi_{i}$
outside the potential: \begin{equation}
\psi_{\mathrm{ref},i}=\sin(k_{i}r+\delta_{i}),\end{equation}
 where $\delta_{i}$ is the phase shift for energy $E_{i}$. For the
reference wave functions, a similar relation to Eq. \prettyref{eq:semi_ortho}
holds \begin{eqnarray}
\left[\psi_{\mathrm{ref},2}\frac{d\psi_{\mathrm{ref},1}}{dr}-\psi_{\mathrm{ref,}1}\frac{d\psi_{\mathrm{ref},2}}{dr}\right]_{0}^{r_{1}}\nonumber \\
=(k_{2}^{2}-k_{1}^{2})\int_{0}^{r_{1}}\psi_{\mathrm{ref},1}\psi_{\mathrm{ref},2}dr.\label{eq:semi_ortho_ref}\end{eqnarray}
 Subtracting Eq. \prettyref{eq:semi_ortho} from Eq. \prettyref{eq:semi_ortho_ref}
leads to 
\begin{eqnarray}
\left(\psi_{\mathrm{ref},1}\frac{d\psi_{\mathrm{ref},2}}{dr}-\psi_{\mathrm{ref},2}\frac{d\psi_{\mathrm{ref},1}}{dr}\right)_{r=0}\qquad\qquad\nonumber \\
\qquad=(k_{2}^{2}-k_{1}^{2})\int_{0}^{r_{1}}dr\,\,(\psi_{\mathrm{ref},1}\psi_{\mathrm{ref},2}-\psi_{1}\psi_{2})\,\,.\label{eq:basic}
\end{eqnarray}
 Then we take $k_{1}=0$ and use the relation 
\begin{equation}
\lim_{k_1\rightarrow 0}k_{1}\cot\delta_{1}=-\frac{1}{a}
\end{equation}
 for the s-wave scattering length $a$ to write 
\begin{equation}
k\cot\delta =  -\frac{1}{a}+\frac{1}{2}k^{2}\rho(k),\label{eq:kcot}
\end{equation}
with
\begin{equation}
\rho(k) \equiv  2\int_{0}^{r_{1}}dr\left(\frac{\psi_{\mathrm{ref},0}}{\sin\delta_{0}}\frac{\psi_{\mathrm{ref}}}{\sin\delta}-\frac{\psi_{0}}{\sin\delta_{0}}\frac{\psi}{\sin\delta}\right),\
\end{equation}
here we use the subscripts zero for the state with $k=0$ and we
omit the subscripts $2$.

 Comparing Eqs. \prettyref{eq:expansion} and \prettyref{eq:kcot},
we get 
\begin{equation}
r_{e}=\rho(k=0)=2\int_{0}^{r_{1}}dr\left(1-\frac{\psi_{0}^{2}}{\sin^{2}\delta_{0}}\right).\label{eq:re}
\end{equation}
 On resonance $(\sin{\delta_0}=1)$, the above equation can be written as 
\begin{equation}
r_{e}=r_{e0}=2\int_{0}^{r_{1}}dr(1-\psi_{0}^{2}),\label{eq:re_res}
\end{equation}
here we introduce $r_{e0}$ denoting the effective range at resonance.

In the case of the square-well potential with depth 
$U_{0}=\hbar^{2}k_{0}^{2}/m$,
$\psi_{0}(r)=\sin{(k_{0}r)}$ for $r<r_{1}$ with $k_{0}r_{1}=(n+\frac{1}{2})\pi$
for ($n=0,1,2,\cdots$) at resonance. Thus we get $r_{e0}=r_{1}$ for
the square-well potential. However, the right hand side of Eq. \prettyref{eq:re_res}
is different from $r_{1}$ for general potentials with a finite range.
As will be seen later, both positive and negative $r_{e0}$ can be obtained 
using a square-well, square-barrier potential.

On resonance ($\delta_0=\pi/2$), the amplitude of the wave function for
$r<r_1$
becomes significant and it can be  comparable to or greater than that
outside the potential.
From Eq. \prettyref{eq:re_res} we see that, while for positive $r_{e0}$
the amplitude inside the potential is suppressed relative to that outside, 
for negative $r_{e0}$ the former is enhanced compared to the latter. 
Although the effect of $r_{e}$ is small away from
resonance due to the first term on the right hand side of Eq.
\prettyref{eq:expansion}
being much larger than the second one, the sign of $r_{e}$ has important
implications for the behavior of the wave function at resonance.

\section{Square-well, square-barrier model
potential\label{sec:sqwb}}

We shall model the microscopic interaction by a single-channel
potential with a square well and a square barrier.  The advantage of
this potential, is that it contains the physics necessary to describe
resonances:  by increasing the thickness of the barrier, or its
height, tunneling through the barrier is decreased, and therefore the
potential can model the effects of changing the strength of coupling
between continuum states and  resonant states in the attractive well.

\subsection{Scattering length and effective range}

\begin{figure}
\begin{center}\includegraphics[%
  width=6cm]{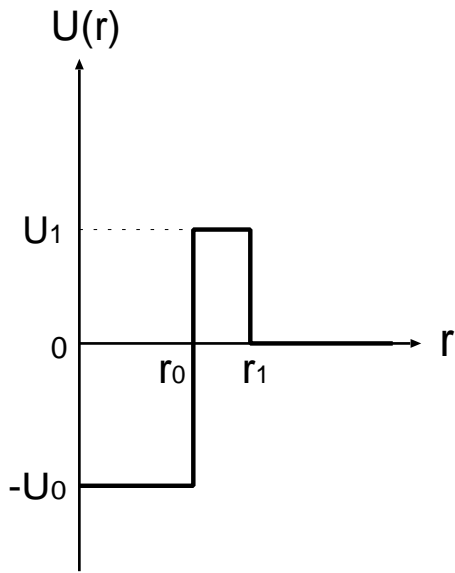}\end{center}
\caption{Schematic picture of the square-well, square-barrier
potential. We take $U_{0},U_{1}>0$ in this paper. \label{fig:pot}}
\end{figure}

Let us consider two atoms interacting through a stepwise potential
with an attractive core of radius $r_{0}>0$ and strength $-U_{0}<0$
and which in addition has a repulsive barrier of width $w=r_{1}-r_{0}$
for $r_{1}>r_{0}$ and a strength $U_{1}>0$ (see Fig. \ref{fig:pot}):
\begin{equation}
U(r)=\left\{ \begin{array}{ccc}
-U_{0} & \mathrm{for} & 0\leq r<r_{0}\,\,,\\
U_{1} & \mathrm{for} & r_{0}\leq r<r_{1}\,\,,\\
\mathrm{0} & \mathrm{for} & r_{1}\leq r<\infty\,\,.\end{array}\right.\label{eq:sqwb}
\end{equation}
 This problem is solved by the standard method of matching the wave
function solutions and their radial derivatives for the above three
regions at
their respective boundaries, $r_{0},$ and $r_{1}$ \cite{Flugge1999}.

The general solutions for the three regions are
\begin{equation}
\psi(r)=\left\{ \begin{array}{ccc}
A\sin(Kr) & \mathrm{for} & 0\leq r< r_{0},\\
B\left(\sinh(\kappa r)+C\cosh(\kappa r)\right) & \mathrm{for} & r_{0}\leq r< r_{1},\\
\sin(kr+\delta) & \mathrm{for} & r_{1}\leq r<\infty.\end{array}\right.\end{equation}
 The well depth parameter is conventionally defined as $k_{0}^{2}=mU_{0}/\hbar^{2}$,
the barrier height parameter as $k_{1}^{2}=mU_{1}/\hbar^{2}$, and
the relative kinetic energy of the two colliding particles is $k^{2}=mE/\hbar^{2}.$
Furthermore, we introduce the auxiliary parameters $K^{2}=m(U_{0}+E)/\hbar^{2}=k_{0}^{2}+k^{2}$,
and $\kappa^{2}=(U_{1}-E)/\hbar^{2}=k_{1}^{2}-k^{2}.$ The energy
dependent s-wave scattering phase shift is denoted by $\delta(k)$.
Equating the logarithmic derivatives at the boundaries, the following
solutions are found for $C$ and $\delta$. First, $C$ is found
from the boundary condition at $r_{0}$ to be 
\begin{equation}
C=\frac{\kappa\tan(Kr_{0})-K\tanh(\kappa r_{0})}{K-\kappa\tan(Kr_{0})\tanh(\kappa r_{0})}\,\,,\label{eq:gamma}
\end{equation}
 and the scattering phase shift is found from the boundary condition
at $r_{1}$
\begin{align}
\delta(k) = & -kr_{1}+\arctan\left(\frac{k}{\kappa}\frac{\tanh(\kappa r_{1})+C}{1+C\tanh(\kappa r_{1})}\right)\nonumber \\
 = & -kr_{1}+\arctan(\mathcal{R}),\label{eq:delta}
\end{align}
where 
\begin{equation}
\mathcal{R}(k)=\frac{k}{\kappa}\frac{\kappa\tan(Kr_{0})+K\tanh[\kappa(r_{1}-r_{0})]}{K+\kappa\tan(Kr_{0})\tanh[\kappa(r_{1}-r_{0})]}.\label{eq:reactance}
\end{equation}
The scattering length is given by
\begin{align}
a = & -\lim_{k\rightarrow0}\frac{\tan\delta(k)}{k}\nonumber \\
  = & r_{1}-\frac{1}{k_{1}\zeta}\left\{ k_{1}\tan(k_{0}r_{0})+k_{0}\tanh[k_{1}(r_{1}-r_{0})]\right\} ,\label{eq:scatt-len}
\end{align}
where 
\begin{equation}
\zeta\equiv k_{0}+k_{1}\tan(k_{0}r_{0})\tanh[k_{1}(r_{1}-r_{0})].\label{eq:scatt-len-denorm}
\end{equation}
 In the present work we are also interested in the effective range,
which can be derived from the low energy expansion \prettyref{eq:expansion}
of the phase shift. Expanding $k\cot\delta$ for Eq. \prettyref{eq:delta}
to the first order in energy, the first term yields the inverse scattering
length as calculated above, and from the second term the following
expression for the effective range is obtained: 
\begin{equation}
r_{e}=r_{1}+r_{\mathrm{r}}+r_{\mathrm{v}},\label{eq:eff-range}
\end{equation}
with 
\begin{align}
r_{\mathrm{r}} = & -\frac{k_{0}^{2}+k_{1}^{2}}{k_{0}k_{1}^{2}(a\zeta)}\left\{ 1+\frac{k_{0}r_{0}}{a\zeta}\mathrm{sech}^{2}[k_{1}(r_{1}-r_{0})]\right\} ,\label{eq:residual}\\
r_{\mathrm{v}} = & \frac{k_{0}^{2}+k_{1}^{2}}{k_{0}k_{1}^{2}(a\zeta)}\frac{r_{1}}{a}\left\{ 1-\frac{1}{k_{1}r_{1}}\tanh[k_{1}(r_{1}-r_{0})]\right\}\nonumber\\
 & +  \frac{1}{k_{1}^{2}a}-\frac{r_{1}^{3}}{3a^{2}}.\label{eq:vanishing}
\end{align}
 At resonance where $a$ diverges (the combination of $a\zeta$ stays
finite), the vanishing part $r_{\rm v}$ is zero,
and the expression for $r_{e}$ reduces to 
\begin{align}
r_{e0} = & r_{1}-\frac{k_{0}^{2}+k_{1}^{2}}{k_{0}^{2}k_{1}}\frac{\tanh[k_{1}(r_{1}-r_{0})]}{1-\tanh^{2}[k_{1}(r_{1}-r_{0})]}\nonumber \\
 & \times \left\{ 1+k_{1}r_{0}\tanh[k_{1}(r_{1}-r_{0})]\right\} ,\label{eq:at-resonance1}
\end{align}
 or equivalently
\begin{equation}
r_{e0}=r_{1}+\frac{k_{0}^{2}+k_{1}^{2}}{k_{0}}\frac{\tan(k_{0}r_{0})-k_{0}r_{0}}{k_{1}^{2}\tan^{2}(k_{0}r_{0})-k_{0}^{2}}.\label{eq:at-resonance2}
\end{equation}
In addition, the condition for resonance ($\zeta=0$) becomes 
$\tan(k_{0}r_{0})<-k_{0}/k_{1}$.
We can see, 
from Eq. \prettyref{eq:at-resonance1}, that the effective range $r_{e0}$
at resonance is smaller than $r_{1}$ since $0\leq\tanh[k_{1}(r_{1}-r_{0})]<1$.
In the weak barrier limit, $\xi\equiv k_{1}(r_{1}-r_{0})\ll1$,
the effective range at resonance can be written as 
\begin{equation}
r_{e0}\simeq r_{0}-\frac{k_{1}}{k_{0}^{2}}\xi+O(\xi^{2}).\label{eq:weak_barrier}\end{equation}
 Thus adding a weak barrier to a square-well potential with the range
$r_{0}$ yields smaller effective range at resonance than that for the original
square-well potential $r_{e0}=r_{0}$. The decrease of $r_{e0}$ means
an enhancement of the amplitude in the potential region, which can
be understood as an effect of blocking probability leakage
due to the additional barrier.

\subsection{Effective range and scattering properties }

\begin{figure}
\begin{center}\includegraphics[%
  width=8cm]{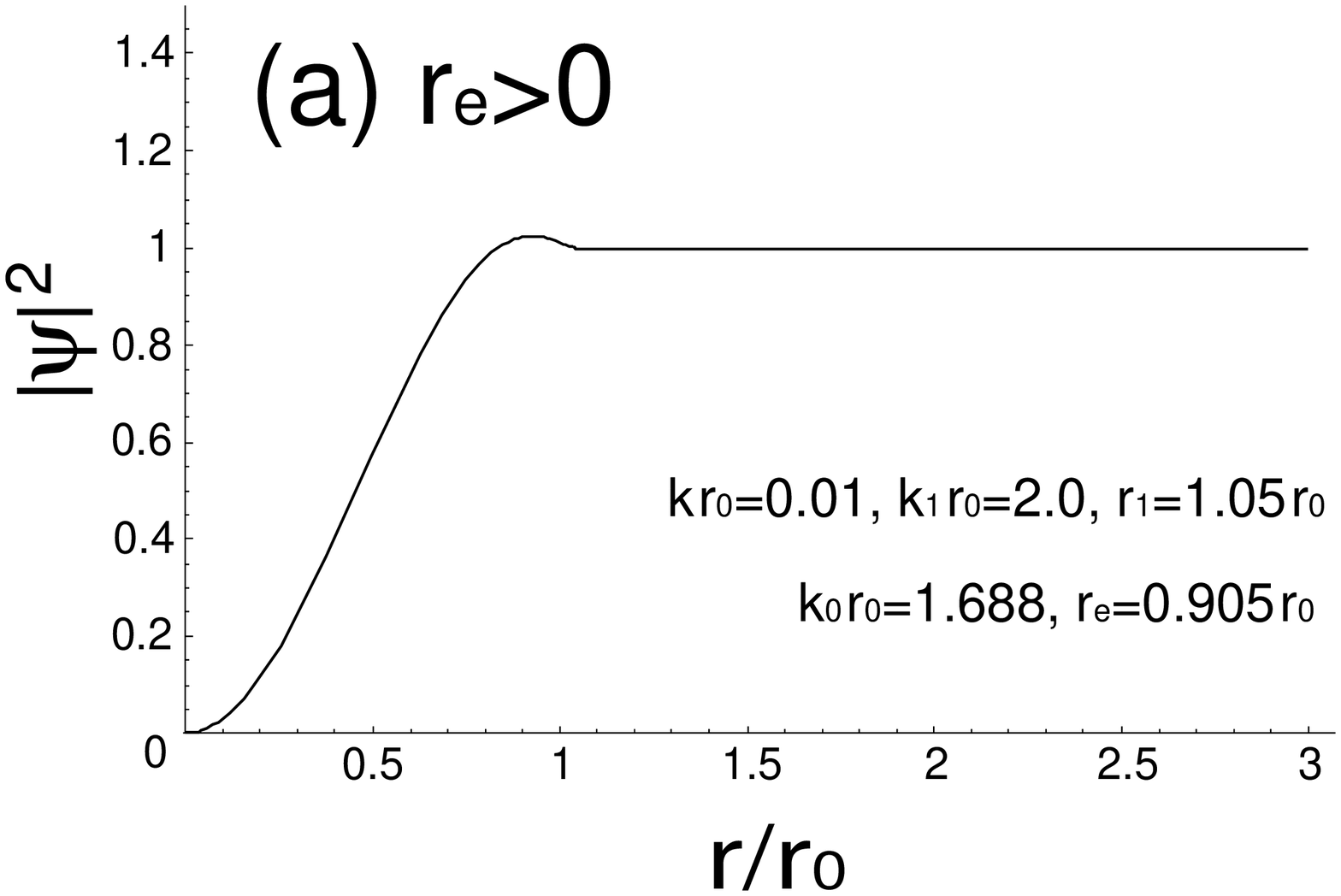}\end{center}
\begin{center}\includegraphics[%
  width=8cm]{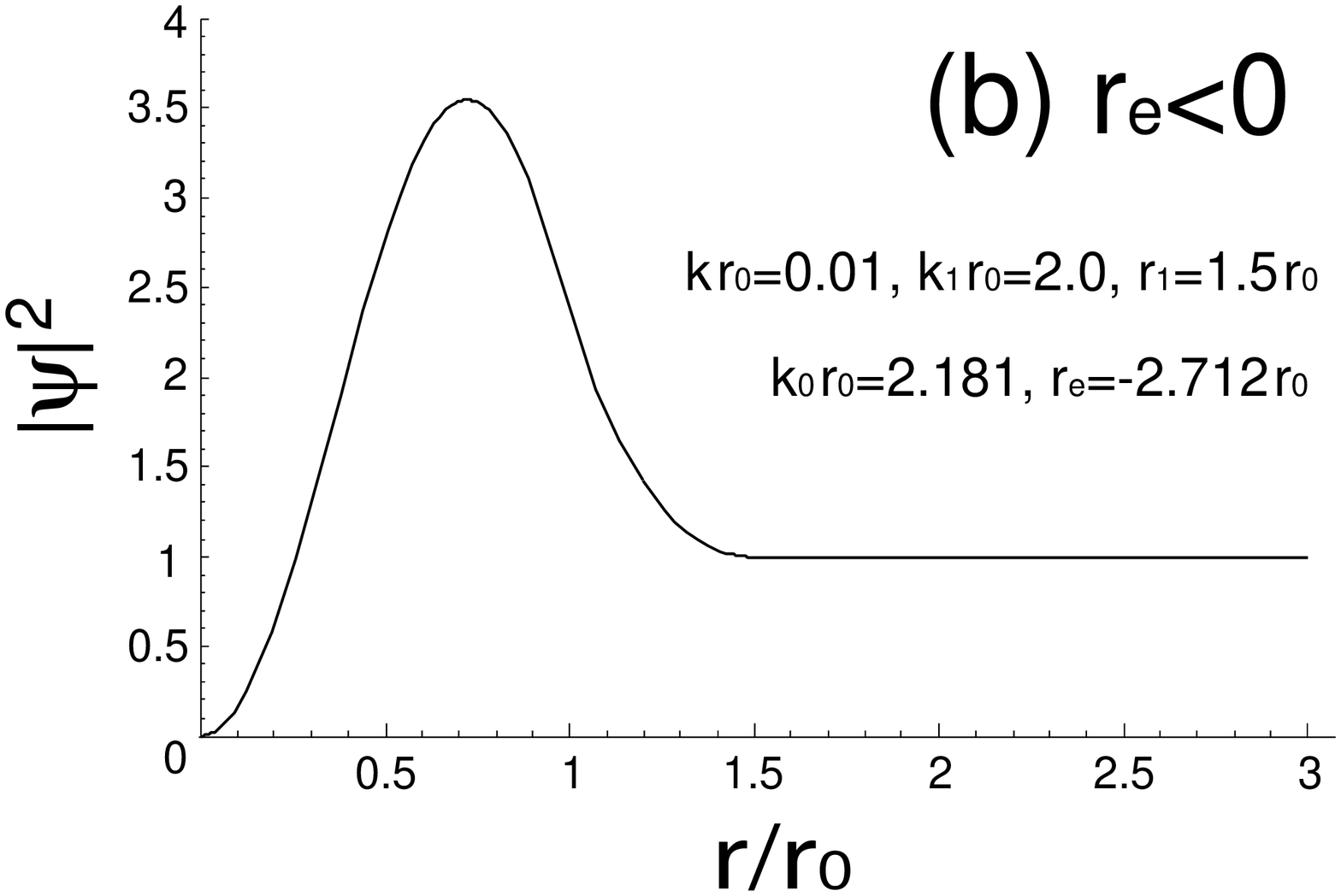}\end{center}
\caption{Resonance wave functions for (a) positive and (b) negative effective
range. In the both cases, we take $k r_0=0.01$ and $k_{1} r_0=2.0$,
and $\psi$ is normalized so that $\psi(r)=\sin(kr+\delta)$ for $r>r_{1}$.
\label{fig:res}}
\end{figure}

To begin with, we illustrate the physical meaning of the effective range
explicitly. In Fig. \ref{fig:res}, we plot the wave functions 
for $|a|\rightarrow \infty$
in the cases of (a) positive and (b) negative effective range for
the same values of $k$, $r_{0}$, $k_{1}$, and different $r_{1}$.
It is clearly seen that the resonance wave function for a negative effective
range has a large amplitude
in the potential range as mentioned before. Even in the case of Fig.
\ref{fig:res} (a), a small bump can be observed close to the edge
of the potential, which means a small decrease of the effective
range compared to the square-well potential 
shown by Eq. \prettyref{eq:weak_barrier}.
At $r_{e}=0$, the mean value of $|\psi|^2$ for $r<r_1$ is unity,
$r_1^{-1}\int_0^{r_1}|\psi|^2 dr=1$, 
thus $\int_{0}^{r_{1}}(1-\psi^{2})dr=0$ as given by Eq. (\ref{eq:re}).

\begin{figure}
\begin{center}\includegraphics[%
  width=8cm]{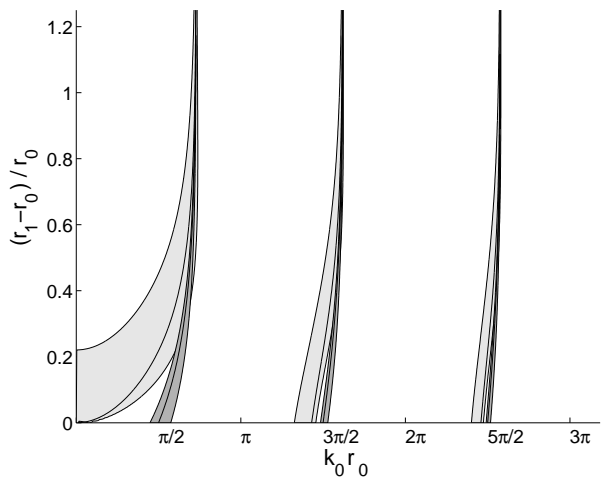}
\end{center}
\caption{Parameter regions of $k_{0}$ and $r_{1}-r_{0}$ at fixed $k_{1}r_0=2$ 
for negative effective range (light gray region) and
for resonance (dark gray region). Here the resonance region is defined
by $|r_0/a|<0.3$; $r_0/a=-0.3$ on the left
boundary of the dark gray area and $r_0/a=0.3$ on that
of the right one. The line plotted in the dark
gray region corresponds to $1/|a|=0$. The boundary of the light gray
region corresponds to
$r_{e}=0$, and the curve in this region shows $r_{e}=-\infty$. \label{fig:condwidth}}
\end{figure}

\begin{figure}
\begin{center}\includegraphics[%
  width=8cm]{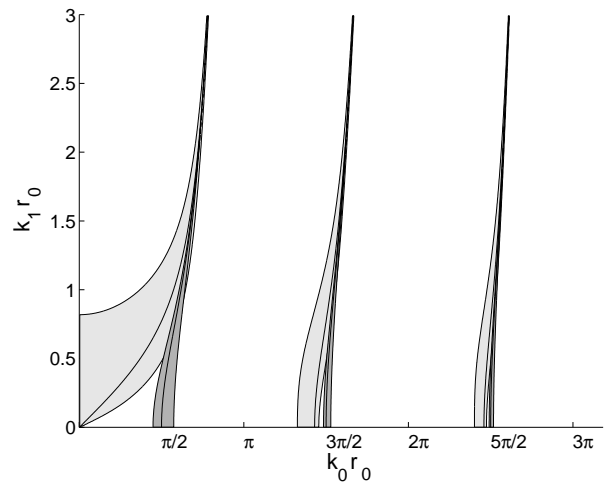}\end{center}
\caption{The same as Fig. \ref{fig:condwidth} for parameter regions of $k_{0}$
and $k_{1}$ at fixed $r_{1}=2r_0$ for negative effective
range (light gray region) and for resonance (dark gray region). \label{fig:condheight}}
\end{figure}

We now discuss the scattering properties
of the square-well, square-barrier potential in detail. In Figs.
\ref{fig:condwidth} and \ref{fig:condheight}, we plot the parameter
regions in which the effective range is negative and the absolute
value of the scattering length is large close to resonance. Figure
\ref{fig:condwidth} shows the dependence of these regions on the
barrier width $r_{1}-r_{0}$ at fixed $r_{0}$ and $k_{1}$. As $r_{1}-r_{0}$
increases, the width of the dark gray regions becomes smaller, which
means the resonance becomes narrower. The effective range at resonance
decreases from $r_{0}$, which is for the square-well potential, to
$-\infty$ with increasing $r_{1}-r_{0}$. This can also be seen from
Eq. (\ref{eq:at-resonance1}) where the absolute value of the second
term increases monotonically from zero to infinity as $r_{1}-r_{0}$
increases with $r_{0}$, $k_{0}$, and $k_{1}$ fixed. Figure \ref{fig:condheight}
shows the dependence of the above regions on the barrier height $k_{1}$
at fixed $r_{0}$ and $r_{1}$. Similarly to the above case, as $k_{1}$
increases, the resonance becomes narrower and $r_{e}$ at resonance
decreases from $r_{0}$ to $-\infty$ monotonically. This continuous
decrease can be proved analytically by 
calculating the derivative of Eq. (\ref{eq:at-resonance2})
with respect to $k_{1}$ and 
by taking the necessary condition for resonance, 
$\tan(k_{0}r_{0})<-k_{0}/k_{1}$, into account.
From both Figs. \ref{fig:condwidth} and \ref{fig:condheight},
we see that the resonance for the deeper potential well (larger $k_{0}$)
is narrower.

\begin{figure}
\begin{center}\includegraphics[%
  width=8cm]{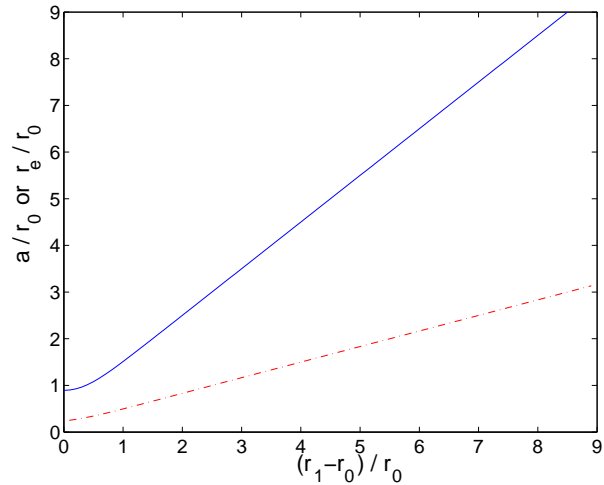}
\end{center}
\caption{(Color online) The scattering length $a$ (solid
line) and the effective range $r_{e}$ (dash-dotted line)
as functions of the barrier width $r_{1}-r_{0}$. 
We take $k_{0}r_0=2$ and $k_{1}r_0=2$. 
\label{fig:off-reswidth}}
\end{figure}

\begin{figure}
\begin{center}\includegraphics[%
  width=8cm]{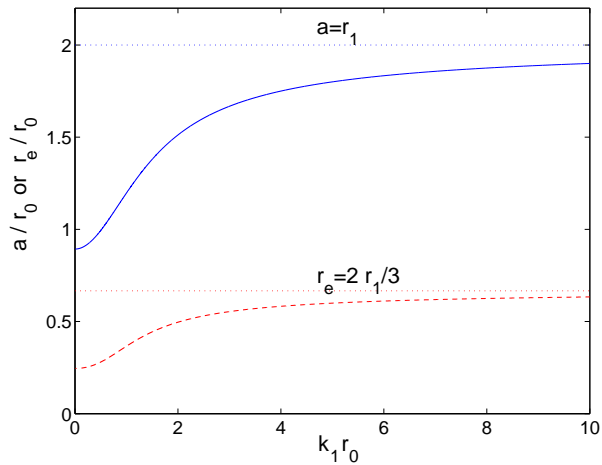}\end{center}
\caption{(Color online) 
The scattering length $a$ (solid
line) and the effective range $r_{e}$ (dashed line)
as functions of the barrier height $k_{1}$. The dotted lines
show the asymptotic values of these quantities. 
We take $k_{0}r_0=2$ and $r_{1}=2r_0$. 
\label{fig:off-resheight}}
\end{figure}

Let us finally mention the behavior of the scattering length and the
effective range away from resonance. In Figs. \ref{fig:off-reswidth}
and \ref{fig:off-resheight}, we plot these quantities as functions
of the width $r_{1}-r_{0}$ and the height parameter $k_{1}$ of the barrier,
respectively. The scattering length, which almost corresponds to the
background one, increases linearly with $r_{1}$ for large barrier widths
(see Fig. \ref{fig:off-reswidth}): \begin{equation}
a\simeq r_{1}-\frac{1}{k_{1}}.\end{equation}
 However, this quantity does not change so much when $k_{1}$ increases.
 It starts from a value comparable to $r_{1}$
at $k_{1}=0$ and converges to $r_{1}$ as $k_{1}\rightarrow\infty$
(see Fig. \ref{fig:off-resheight}).   Whether it approaches $r_{1}$ 
from above or below depends on
the value of $k_{0}r_{0}$.   The behavior of the effective
range is similar to that of the scattering length.
It increases linearly with $r_{1}$ at large $r_{1}-r_{0}$
(Fig. \ref{fig:off-reswidth}):
\begin{equation} r_{e}\sim\frac{2}{3}r_{1},\end{equation}
 and, with increasing $k_{1}$, it remains of order $r_{1}$
and converges to $2r_{1}/3$ as $k_{1}\rightarrow\infty$ 
(Fig. \ref{fig:off-resheight}).

\section{non-universal zero temperature BCS-BEC crossover \label{sec:crossover} }

In the present section we examine the BCS-BEC crossover problem for a
gas of equal numbers $N/2$ in two different internal states interacting 
through the square-well, square-barrier potential within an effective 
single-channel model.

\subsection{Basic formalism}
The Hamiltonian
$H$ for the two-component Fermi gas is  
\begin{align}
H  = & \sum_{\vec{k},\sigma}\varepsilon_{\vec{k}}a_{\vec{k}\sigma}^{\dagger}a_{\vec{k}\sigma}\nonumber \\
 & +  \frac{1}{\mathcal{V}}\sum_{\vec{k}\vec{k}^{\prime}}U_{\vec{k}\vec{k}^{\prime}}a_{\vec{k}\uparrow}^{\dagger}a_{-\vec{k}\downarrow}^{\dagger}a_{-\vec{k}^{\prime}\downarrow}a_{\vec{k}\uparrow},
\end{align}
where $a_{\vec{k}\sigma}^{\dagger},a_{\vec{k}\sigma}$ are the creation
and annihilation operators, respectively, of atoms with wave vector
$\vec{k},$ and pseudospin index, $\sigma=\ \uparrow\mathrm{or}\downarrow.$
The single particle kinetic energy is $\varepsilon_{\vec{k}}=\hbar^{2}\vec{k}^{2}/(2m),$
and the inter-atomic interaction in momentum space is $U_{\vec{k}\vec{k}^{\prime}}.$
The zero temperature BCS-BEC crossover problem is here numerically
studied along the lines of Refs. 
\cite{Leggett1980,Diener2004a,Diener2004b,Palo2004,Parish2005a}
based on the BCS variational state
\begin{equation}
|\mathrm{\Psi_{\mathrm{BCS}}}\rangle=\Pi_{\vec{k}}\left(u_{\vec{k}}+v_{\vec{k}}a_{\vec{k}\uparrow}^{\dagger}a_{-\vec{k}\downarrow}^{\dagger}\right)|\Psi_{0}\rangle,
\end{equation}
where $|\Psi_{0}\rangle$ is the vacuum state, and $u_{\vec{k}}$
and $v_{\vec{k}}$ are the variational parameters normalized 
according to the condition $|u_{\vec{k}}|^{2}+|v_{\vec{k}}|^{2}=1$.
The energy of the BCS state $E(u_{\vec{k}},v_{\vec{k}})=\langle\Psi_{\mathrm{BCS}}|H|\Psi_{\mathrm{BCS}}\rangle$
is to be minimized with respect to the variational parameters
$u_{\vec{k}}$ and $v_{\vec{k}}$, subject to the normalization condition 
and the requirement that the mean particle number
$N(u_{\vec{k}},v_{\vec{k}})
=\sum_{\vec{k},\sigma}\langle\Psi_{\mathrm{BCS}}|a_{\vec{k}\sigma}^{\dagger}a_{\vec{k}\sigma}|\Psi_{\mathrm{BCS}}\rangle
=2\sum_{\vec{k}}|v_{\vec{k}}|^{2}$ should be fixed.
By introducing the Lagrange multipliers $E_{\vec{k}}$ and $\mu$
to ensure normalization and number conservation, the stationary condition 
$\delta[E-\mu N-\sum_{\vec{k}}E_{\vec{k}}(|u_{\vec{k}}|^{2}+|v_{\vec{k}}|^{2})]/\delta(u_{\vec{k}},v_{\vec{k}})=\vec{0}$ 
yields
$\xi_{\vec{k}}u_{\vec{k}}+\Delta_{\vec{k}}v_{\vec{k}}=E_{\vec{k}}u_{\vec{k}}$
and $u_{\vec{k}}\Delta_{\vec{k}}^{\ast}-\xi_{\vec{k}}v_{\vec{k}}=E_{\vec{k}}v_{\vec{k}}$, where $\xi_{\vec{k}}\equiv \hbar^2k^2/(2m) -\mu$.
Here, the momentum dependent gap function is defined as 
$\Delta_{\vec{k}}=\sum_{\vec{k}^{\prime}}U_{\vec{k}\vec{k}^{\prime}}u_{\vec{k}^{\prime}}v_{\vec{k}^{\prime}}^{\ast}$.
By solving the above equations, one determines the usual quasi-particle
energy $E_{\vec{k}}=(\xi_{\vec{k}}^{2}+\Delta_{\vec{k}}^{2})^{1/2}$
and the coherence factors $|u_{\vec{k}}|^{2}=(1+\xi_{\vec{k}}/E_{\vec{k}})/2$
and $|v_{\vec{k}}|^{2}=(1-\xi_{\vec{k}}/E_{\vec{k}})/2$.
Inserting the expression for
$u_{\vec{k}}v_{\vec{k}}^{\ast}=\Delta_{\vec{k}}/(2E_{\vec{k}})$
into the definition of the gap yields the gap equation:
\begin{equation}
\Delta_{\vec{k}}=-\frac{1}{\mathcal{V}}\sum_{\vec{k}^{\prime}}U_{\vec{k}\vec{k}^{\prime}}\frac{\Delta_{\vec{k}^{\prime}}}{2E_{\vec{k}^{\prime}}}.\label{eq:gap-eq}
\end{equation}
This is to be solved self-consistently together with the equation
for the atomic number density 
\begin{equation}
n=\frac{1}{\mathcal{V}}\sum_{\vec{k}}\left(1-\frac{\xi_{\vec{k}}}{E_{\vec{k}}}\right).\label{eq:number-eq}
\end{equation}
At fixed density $n=k_{\mathrm{F}}^{3}/(3\pi^{2})$, where
$k_{\mathrm{F}}$ is the Fermi momentum for the non-interacting gas, 
Eq. \prettyref{eq:number-eq}
determines the chemical potential, $\mu$. For the contact
interaction, the regularized homogeneous BCS-BEC crossover problem
\cite{Eagles1969b,Leggett1980} is analytically solvable
\cite{Haussmann1993,Marini1998,Papenbrock1999}
and results in the zero range (Leggett) reference curves in Figs. 
\ref{cap:reffsqwb}, \ref{cap:mu}, \ref{cap:gap-at-zero}, \ref{cap:nkhalf},
and \ref{cap:cfsqwb}.

The partial wave decomposition of the interaction potential is
\begin{align}
U_{\vec{k}\vec{k}^{\prime}} = & \int d^{3}r\  e^{i(\vec{k}-\vec{k}^{\prime})\cdot\vec{r}}\ U(\vec{r})\nonumber \\
  = & \sum_{\ell=0}^{\infty}(2\ell+1)4\pi U_{\ell}(k,k^{\prime})P_{\ell}(\hat{k}\cdot\hat{k}^{\prime}),
\end{align}
with 
\begin{equation}
U_{\ell}(k,k^{\prime})=\int_{0}^{\infty}dr\  r^{2}j_{\ell}(kr)U(r)j_{\ell}(k^{\prime}r).
\end{equation}
Here $j_{\ell}(x)=\sqrt{\pi/(2x)}J_{\ell+1/2}(x)$ is the $\ell$'th
spherical Bessel function defined from the $\nu$'th ordinary Bessel
function of the first kind, $J_{\nu}(x),$ and $P_{\ell}$ is
the $\ell$'th Legendre function with $\hat{k}\cdot\hat{k}^{\prime}=\cos\theta$,
$\theta$ being the angle between the unit vectors 
$\hat{k}$ and $\hat{k}^{\prime}$.
We also used the addition formula 
$(2\ell+1)P_{\ell}(\hat{k}\cdot\hat{k}^{\prime})=4\pi\sum_{m=-\ell}^{\ell}Y_{\ell m}^{\ast}(\hat{k})Y_{\ell m}(\hat{k}^{\prime})$
for the spherical harmonic functions $Y_{\ell m}(\hat{k})$. 
The momentum dependent gap can be decomposed in the same way as
\begin{equation}
\Delta_{\vec{k}}=\sum_{\ell=0}^{\infty}\sum_{m=-\ell}^{m}\Delta_{\ell m}j_{\ell}(k)Y_{\ell m}(\hat{k}),
\end{equation}
where the $\Delta_{\ell m}$'s are the angular weight coefficients.
With the assumption that the main low energy contribution to the gap is of
s-wave $(\ell=0)$ character, the gap equation reduces to 
\begin{equation}
\Delta(k)=-\frac{1}{\pi}\int dk^{\prime}\  k^{\prime2}\frac{U_{0}(k,k^{\prime})\Delta(k^{\prime})}{\sqrt{\xi_{k^{\prime}}^{2}+\Delta^{2}(k^{\prime})}},\label{eq:1d-gap-eq}
\end{equation}
where the s-wave part of the potential is 
\begin{align}
U_{0}(k,k^{\prime}) = & \int_{0}^{\infty}dr\  r^{2}j_{0}(kr)U(r)j_{0}(k^{\prime}r)\nonumber \\
 = & \frac{U_0+U_1}{2kk'}\left[\frac{\sin{(|k+k'|r_0)}}{|k+k'|}-\frac{\sin{(|k-k'|r_0)}}{|k-k'|}\right]\nonumber\\
 & - \frac{U_1}{2kk'}\left[\frac{\sin{(|k+k'|r_1)}}{|k+k'|}-\frac{\sin{(|k-k'|r_1)}}{|k-k'|}\right].\label{eq:Ukk}
\end{align}

\subsection{Solution of the crossover equations}

The crossover equations consist of Eq. \prettyref{eq:number-eq}
and Eq. \prettyref{eq:1d-gap-eq} together with the expression 
\prettyref{eq:Ukk} for the momentum dependent interaction. 
We notice that, for a finite range interaction potential vanishing
beyond a scale $r_{1}$, all momentum integrals are cutoff typically
at the scale of $1/r_{1}$ and therefore remain ultraviolet convergent.
Here, $r_{1}$ characterizes the range of the interatomic interactions.
In the context of ultracold atoms,
$r_{1}$ can be taken to be the characteristic scale of the van der
Waals interaction. We shall consider a system at fixed density
and will study the properties of the system as the potential is changed
by varying the well depth, the barrier height or the barrier width.
Finally, in order to describe the crossover in terms of physically relevant
two-body quantities, we parametrically plot quantities
versus the scattering length $a$ obtained from Eq. \prettyref{eq:scatt-len}
and the effective range $r_{e}$ obtained from Eq. \prettyref{eq:eff-range}
as functions of the model parameters.

Below, we briefly describe our numerical procedure used to solve the
crossover equations. By discretizing the momenta on a finite grid,
the gap equation \prettyref{eq:1d-gap-eq} and the number density
equation \prettyref{eq:number-eq} were solved iteratively 
for the given density.
The spacing of the momentum grid, $\Delta k$, 
was chosen to provide sufficient sampling on the scale
of $k_{\mathrm{F}}\sim n^{1/3}$.
The maximum momentum $k_{\rm max}$
was chosen to be much larger than 
the scales of the internal range of the potential 
$r_{0}^{-1}$ and $r_{1}^{-1}$.
All results shown below are calculated with $\Delta k=k_{\rm F}/15$
and $k_{\rm max}=2500 \Delta k$ (this corresponds to 
$k_{\rm max}\simeq 50 r_0^{-1}$ for the model parameters employed).
The system was driven from the BEC to the BCS limit and the initial
guess for the $k$-dependent gap function $\Delta(k)$ was determined from the
Fourier transform of the real space wave function determined in 
Section \ref{sec:sqwb}. The equations were solved by iteration,
and the process was continued until the difference in $\mu$ between successive
iterations was less than one part in $10^6$.
The results were checked to be insensitive to further increase of
grid size, $k_{\rm max}$, and decrease of grid spacing, $\Delta k$.

\begin{figure}
\begin{center}\includegraphics[%
  width=8cm,
  keepaspectratio]{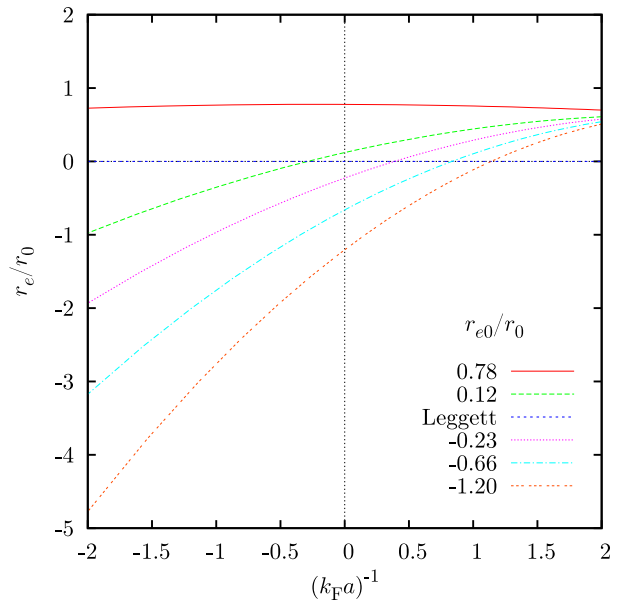}\end{center}
\caption{(Color online) The effective range $r_{e}/r_{0}$ as a function
of the dimensionless constant $(k_{F}a)^{-1}$. 
The density is set to be $k_{\rm F}r_0\simeq 0.31$.
\label{cap:reffsqwb}}
\end{figure}

We consider the case of driving the BCS-BEC crossover by varying the
potential well parameter $k_{0}$ around its critical value $k_{0c}$ at
which the scattering length diverges.  Both the scattering length and
the effective range depend on $k_{0}$ and therefore $r_{e}/r_{0}$
depends on $(k_{\mathrm{F}}a)^{-1}$ through the parameter $k_{0}$. Figure
\ref{cap:reffsqwb} contains parametric plots of the effective range
versus the scattering length for different values 
of the barrier width $r_{1}-r_{0}$.
As we are interested in the scattering properties of the model and not
in the particular values of the model parameters, it is more natural to
show the $r_{e}$ curves in terms of the effective range $r_{e0}$ at
$(k_{\mathrm{F}}a)^{-1}=0$. The barrier widths used in the calculations,
$(r_{1}-r_{0})/r_0=0.1,0.25,0.30,0.35$, and $0.40$, 
correspond to effective ranges
$r_{e0}/r_{0}\approx0.78,0.12,-0.23,-0.66$, and -1.20.  The remaining
parameter $k_1$ was chosen to be $k_{1}r_0=2.0$, and the density
was chosen as $n=0.001r_0^{-3}$, i.e., $k_{\rm F}r_0\simeq 0.31$.
For relative
large and positive $r_{e0}$, the effective range is only weakly dependent
on the coupling constant in the crossover region as evidenced by the solid
line in Fig. \ref{cap:reffsqwb}. As $r_{e0}$ decreases, the effective
range develops a strong dependence on the inverse coupling constant.
We note that
similar results for $r_{e}$ presented in Fig. \ref{cap:reffsqwb} could
have been obtained by changing the barrier height instead.  Such a
degeneracy is to be expected for potentials with many model parameters
since the mapping of the model parameters to the effective low
energy parameters is a many-to-two mapping.  In the Eagles-Leggett
model, only the scattering length appears and it is therefore interpreted
as the $r_{e0}=0$ case with $r_{e}=0$ for all couplings. 

\begin{figure}
\begin{center}\includegraphics[%
  width=8cm,
  keepaspectratio]{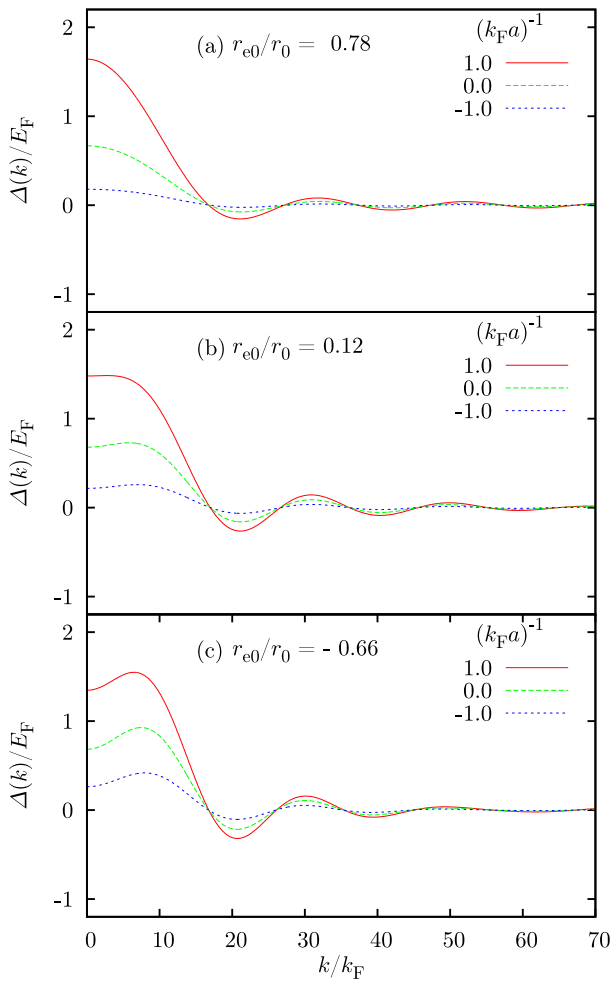}\vspace{0.05cm}\end{center}
\caption{(Color online) The solution of the gap equation for the values $(k_{\mathrm{F}}a)^{-1}=-1,0$, and 1
of the inverse coupling constant is presented for several values of
the effective range $r_{e0}$ in the unitarity limit. For positive effective
range (a), the zero momentum gap is a local maximum. Close to zero
effective range, the maximum turns into a plateau with a vanishing
momentum gradient. In (c), the zero momentum gap is turned into a local
minimum for all values of the inverse coupling. 
The density is set to be $k_{\rm F}r_0\simeq 0.31$.
\label{cap:gap3sqwb}}
\end{figure}

\subsection{Energy gap}

We first examine the effective range dependence of
the gap parameter $\Delta(k)$.
In panels (a)-(c) of Fig. \ref{cap:gap3sqwb}, $\Delta(k)/E_{\rm F}$
obtained from the crossover equations is displayed 
for several values of $r_{e0}$, here $E_{\rm F}\equiv\hbar^2 k_{\rm F}/2m$ 
is the Fermi energy for the non-interacting gas.
The solid curves in panels (a)-(c) show the evolution
of $\Delta(k)$ in the near BCS limit. 
In this limit, the gap has a maximum at $k=0$ for 
relatively large and positive $r_{e0}$ as displayed in panel (a).
As $r_{e0}$ decreases, the maximum flattens and the sign of
the curvature at $k=0$ changes from negative to positive
as can be seen by comparing the solid curves in panels (a) and (b).
For negative $r_{e0}$, the gap has a local minimum at $k=0$ 
and the maximum of the gap moves to a higher $k$
[see the solid curve in panel (c)].

The energy gap in the ground state is the minimum of the quasiparticle
dispersion
$E_{\vec{k}}$ for all $\vec{k}$ \cite{Leggett1980} and, for a Fermi
gas with contact interactions, the gap $\Delta_{\vec{k}}=\Delta$ is
momentum independent. For a positive $\mu$, 
the energy $E_{\vec{k}}=(\xi_{\vec{k}}^{2}+\Delta^{2})^{1/2}$
is minimized at $\hbar^{2}k^{2}/2m=\mu$ 
with the value $E_{\mathrm{min}}=|\Delta|$.
On the other hand, for a negative $\mu$, 
the energy minimum $E_{\mathrm{min}}=(\mu^{2}+\Delta^{2})^{1/2}$
is attained at $k=0$. Therefore, as the interaction is decreased
from the BEC limit to the BCS one, the $k_{\mathrm{min}}$ value, where
$E_{\vec{k}}$ has its minimum, shifts from $0$ to $\sqrt{2m\mu/\hbar^{2}}$
at $\mu=0$ to $k_{\mathrm{min}}=k_{\rm F}$ in the BCS limit. 
Due to the interaction-induced shift in the position
of $k_{\mathrm{min}}$, the functional behavior of $E_{\vec{k}}$ is
also directly reflected in the pair wave function $\psi(k)=\Delta/(2E_{\vec{k}})$
which appears directly in the gap equation, and also in a number of
derived quantities such as the healing length and condensate fraction.
In the BEC limit, where the gap equation reduces to the equation
for two bound particles, 
$\psi(k)$ represents the internal wave function
of the pair. The $1/E_{\vec{k}}$ dependence implies a maximum of
$\psi(k)$ at $k=0$ for $\mu<0$ (the BEC case) and at
$\sqrt{2m\mu}/\hbar$
for $\mu>0$, the BCS case (see also \cite{Parish2005a}).

Next we consider the functional behavior in the general
case of a momentum dependent gap $\Delta_{\vec{k}}$.  We assume that the
Fermi gas  is so dilute that the range
of the potential is much shorter than the interatomic spacing, i.e.,
$r_{0},r_{1}\ll k_{\mathrm{F}}^{-1}$.
The momentum dependent gap varies on the scale of $1/r_{0}$ which is
large compared with $k_{\mathrm{F}}$, and therefore the behaviors
described above for a constant gap are expected to be rather general
\cite{note-1}.

\begin{figure}
\begin{center}\includegraphics[%
  width=8cm,
  keepaspectratio]{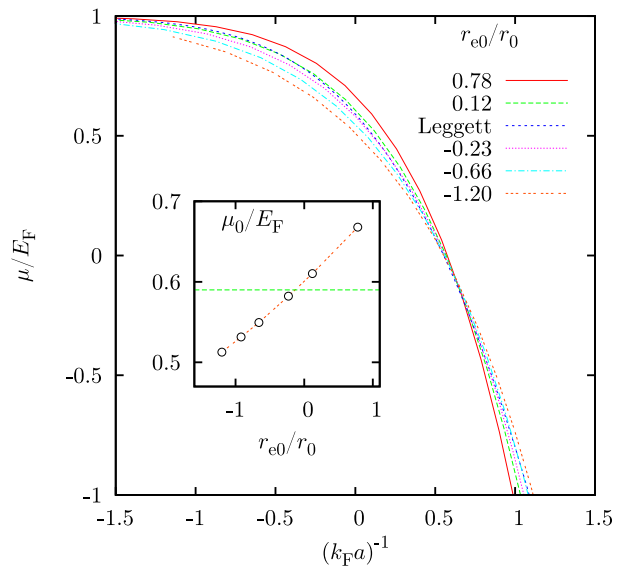}\end{center}
\caption{(Color online) The chemical potential $\mu/E_{\mathrm{F}}$ versus
$(k_{\mathrm{F}}a)^{-1}$ is plotted for changing the depth of the
potential well \cite{note-2}.
Each curve is labeled by the value
of the dimensionless effective range $r_{e0}/r_0$ in the
unitarity limit. As the effective range becomes negative,
the finite range correction to the universal chemical potential changes its
sign. In the inset, we have plotted the chemical potential $\mu_0$
in the unitarity limit, $(k_{\mathrm{F}}a)^{-1}=0$, 
as a function of the unitary effective range $r_{e0}$. 
The universal result $\mu_0/E_{\mathrm{F}}\approx 0.59$
obtained by Leggett \cite{Leggett1980} is marked with the 
horizontal dashed line. 
The density is set to be $k_{\rm F}r_0\simeq 0.31$.
\label{cap:mu}
}
\end{figure}

\subsection{Chemical potential}

Next we discuss the influence of the nonzero effective
range on the chemical potential. Figure \ref{cap:mu} contains a plot
of the chemical potential as a function of $(k_{\rm F}a)^{-1}$ for
several values of $r_{e0}$. At a sufficiently low value of $r_{e0}$,
the chemical potential approaches the universal result (short blue
dashes denoted by ``Leggett'') as obtained from the zero-range model. 
By an additional variation
of the barrier parameter, the effective range $r_{e0}/r_{0}$ is varied
from a positive value $0.78$ (solid line) to $-1.20$ (dotted line).
The plots of the chemical potential for different values of $r_{e0}$
are all seen to cross at $(k_{\mathrm{F}}a)^{-1}\approx 0.55$,
where $\mu$ is close to zero.
The inset in Fig. \ref{cap:mu} contains a plot of the chemical potential 
$\mu_0$ at $k_{\mathrm{F}}|a|\rightarrow \infty$ 
as a function of the effective range, which clearly shows 
the correction due to the effective range. 
The horizontal dashed line of $\mu_0/E_{\rm F}\approx 0.59$ corresponds
to $\mu_0$ for a contact potential, and the crossing of this line
at $r_{e0}\approx0$ also validates our approach
to discuss non-universal corrections by use of a finite-range model
potential.  For positive $r_{e0}$, the chemical potential for
$k_{\mathrm{F}}|a|\rightarrow \infty$
is enhanced with respect to the zero range case as was also observed in
Ref. \cite{Parish2005a} based on the Gaussian potential.
On the other hand for a negative $r_{e0}$
the chemical potential is 
reduced compared to the curve for a contact potential. 
These non-universal corrections become small with decreasing density.
We have observed that, for a lower density of $n=0.0001r_0^{-3}$,
i.e., $k_{\rm F}r_0\simeq 0.14$, deviations from the Leggett curve
in the region $-1\alt (k_{\rm F}a)^{-1} \alt 1$ are about a factor of 
two smaller than those in Fig. \ref{cap:mu}.

By matching the scattering properties of single- and two-channel
models on resonance, it is found that $4\pi\hbar^{2}a/m=-g^{2}/\nu$
and $r_{e0}=-8\pi\hbar^{4}/(mg^{2})$, where $g$ is the atom-molecule
coupling and $\nu$ is the detuning in the two-channel boson-fermion
model \cite{Bruun2004b}. Because of $r_{e0}$ is negative in this model,
for $(k_{\rm F}a)^{-1}\simeq0$,
the above non-universal reduction of the chemical potential is expected in less
wide to narrow Feshbach resonances \cite{Palo2004,Diener2004b}.
It is also noted that this reduction close to a narrower resonance
cannot be achieved with a monotonic attractive potential 
(which all have $r_{e0}>0$) and therefore
constrains the model potential when modelling atom gases.

\begin{figure}
\begin{center}\includegraphics[%
  width=8cm]{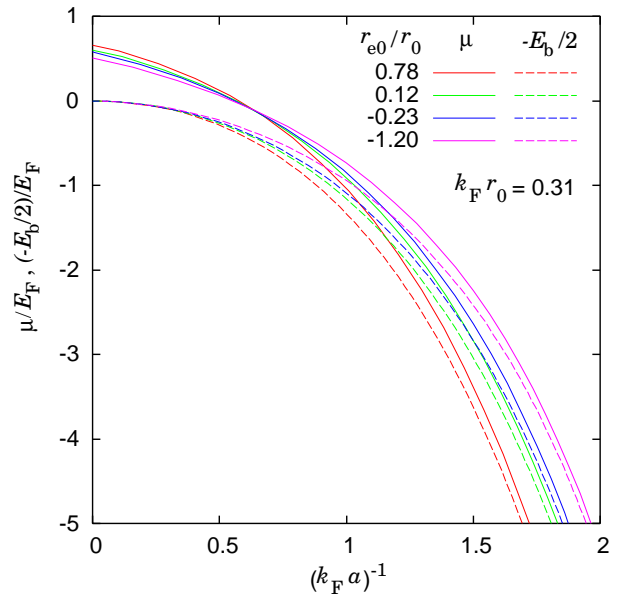}\end{center}
\caption{(Color online) Comparison between the chemical potential (solid lines)
and the binding energy $E_{\mathrm{b}}$ of the two-body bound state
(dashed lines) for several values of the
effective range $r_{e0}/r_{0}=0.78$ (red), $0.12$ (green), $-0.23$ (blue),
and $\ -1.2$ (magenta).
The density is set to be $k_{\rm F}r_0\simeq 0.31$.
\label{cap:energy}}
\end{figure}

In Fig. \ref{cap:energy}, we compare the chemical potential and 
the binding energy $E_{\mathrm{b}}$ of the two-body bound state
of the same square-well, square-barrier potential.
For $a\gg r_0, r_1$, the binding energy is well approximated by
using only $a$ and $r_e$ as $E_{\rm b}\simeq (\hbar^2/ma^2) (1+r_e/a)$
($E_{\rm b}$ is defined to be positive),
and thus the negative effective range reduces the binding energy
close to threshold.  For $a\alt r_0, r_1$, 
since the extension of the wave function of the bound state is comparable to
the range of the potential, $E_{\rm b}$ depends strongly on 
its specific shape. Therefore, $E_{\rm b}$ for a potential
with a larger negative $r_{e0}$, which has a stronger barrier
[i.e., larger $\xi= k_{1}(r_{1}-r_{0})$], is smaller 
at $(k_{\rm F}a)^{-1}\gg 1$.
We can also see that the chemical potential consistently behaves
as $\mu\to-E_{\mathrm{b}}/2$ in this limit.

\subsection{Consequences of the nonzero effective range in non-universal resonances}

Most experiments on the BCS-BEC crossover done so far have been performed with
broad resonances. According to an estimate in \cite{Diener2004b},
$k_{\mathrm{F}}r_e\sim 0.01$ and $r_e/r_0\alt 1$
\footnote[3]{Comparing the real part of the T-matrix on resonance
in their two-channel model, and to the present single-channel model
with the effective range expansion, we observe that $r^*$ in 
\cite{Diener2004b} corresponds to $|r_e|/4$.} 
in the experiment by Regal {\it et al.} \cite{Regal2004}, and 
$k_{\rm F}r_e\sim 10^{-4}$ and $r_e/r_0\sim 0.1$ in the experiment by
Zwierlein {\it et al.} \cite{Zwierlein2004a}.
Due to the small values of $k_{\rm F}r_e$, the effective range
has an imperceptible effect in these experiments.
However, it is essential to study narrower resonances, in which
effective range corrections are significant,
for understanding strongly interacting Fermi gases more deeply.
Therefore, consequences of the nonzero effective range, especially for 
negative cases, in future experiments on narrow resonances are 
worth discussing.
In the remaining part of this paper, we study effective range corrections
for uniform systems on the following four measurable quantities.
zero momentum gap, momentum distribution, and condensate fraction.

\subsubsection{Zero momentum gap\label{sec:gap-at-zero}}

\begin{figure}
\begin{center}\includegraphics[%
  width=8cm,
  keepaspectratio]{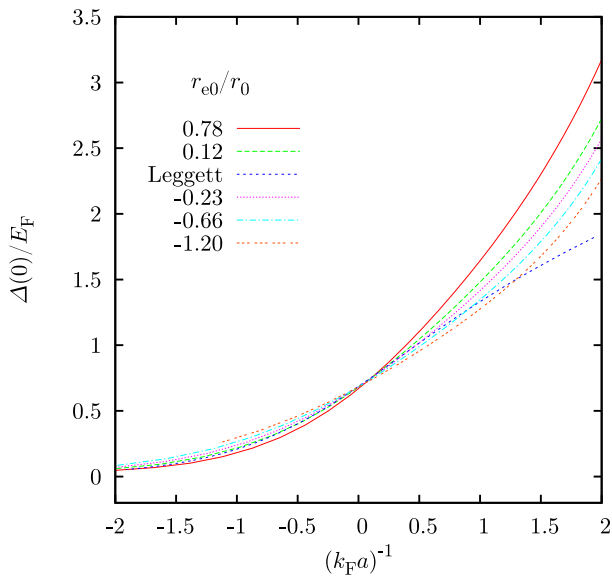}\end{center}
\caption{(Color online) The zero temperature gap $\Delta(0)/E_{\mathrm{F}}$
at $k=0$ is plotted as a function of the inverse coupling constant
$(k_{\mathrm{F}}a)^{-1}$ for several values of the effective range
$r_{\mathrm{e0}}/r_{0}$. 
The density is set to be $k_{\rm F}r_0\simeq 0.31$.
\label{cap:gap-at-zero} }
\end{figure}

In Fig. \ref{cap:gap-at-zero} the zero momentum
gap $\Delta(0)$ is plotted as a function of $(k_{\mathrm{F}}a)^{-1}$. 
All curves cross at almost the same point near $(k_{\rm F}a)^{-1}=0$
and thus the zero momentum gap shows little dependence
on $r_{e0}$ close to resonance.
In the BCS region [$(k_{\rm F}a)^{-1}<0$], however, 
the effective range corrections can be significant
and $\Delta(0)$ is enhanced for large and negative $r_{e0}$.
A larger negative value of $r_{e0}$, which is yielded by a stronger barrier
[larger $\xi= k_{1}(r_{1}-r_{0})$] in the present single-channel model,
corresponds to a weaker atom-molecule coupling $g$ in a two-channel model.
Thus, in this case, BCS pairing is more favorable on the BCS side 
and the gap is enhanced.
This result can be also understood in terms of the effective range expansion
introducing the effective scattering length $a_{e}(k)$:
\begin{equation}
  \frac{1}{a_e(k)}\equiv \frac{1}{a}-\frac{1}{2}r_e k^2.
\label{ae}
\end{equation}
For negative $a$ and $r_e$, the absolute value of 
the effective scattering length $|a_e|$
is larger than $|a|$; i.e., the effective range correction 
due to a negative $r_e$ yields a stronger attractive interaction
on the BCS side, which leads to a larger BCS gap.
Using the effective scattering length, we may write
$\Delta(0)/E_{\mathrm{F}}\approx(8/e^{2})e^{-2\{k_{\mathrm{F}}a_e(k_{\rm F})\}^{-1}/\pi}
=(8/e^2)e^{\{2(k_{\rm F}a)^{-1}-k_{\rm F}r_e\}/\pi}$ in the BCS limit,
which converges toward $(8/e^{2})e^{-2(k_{\mathrm{F}}a)^{-1}/\pi}$
in the far BCS limit.

In the far BEC regime [$(k_{\rm F}a)^{-1}\gg 1$], where the molecules 
are deeply bound, $\Delta(0)$ also strongly depends on 
the specific properties of the potential, as has been seen 
for the chemical potential,
and therefore it deviates from the zero range result.
The suppression of $\Delta(0)$ for a large, negative $r_{e0}$ compared to
that for a small negative or positive $r_{e0}$ can be attributed
to the stronger repulsive barrier in the former case,
which acts to destroy the pair in this regime.
We have also confirmed that,  with decreasing density,
$\Delta(0)$ for each value of $r_{e0}$ approaches
the one for a contact potential. 
For a lower density of $n=0.0001r_0^{-3}$, the effective range corrections
in the region of $-1\alt (k_{\rm F}a)^{-1} \alt 1$ are about a factor of 
two smaller than those in Fig. \ref{cap:gap-at-zero},
as in the case of the chemical potential
\footnote[4]{Also for the momentum distribution and the condensate fraction 
discussed later, the effective range corrections in the region of
$-1\alt (k_{\rm F}a)^{-1} \alt 1$ at $n=0.0001r_0^{-3}$
($k_{\rm F}r_0\simeq0.14$) are about a factor of two smaller than
those are at $n=0.001r_0^{-3}$ ($k_{\rm F}r_0\simeq0.31$).
This fact suggests that the effective range correction 
scales roughly as $\sim k_{\rm F}|r_{e0}|$ under density variations.
}.

\subsubsection{Momentum distribution}

\begin{figure}
\begin{center}\includegraphics[%
  width=8cm,
  keepaspectratio]{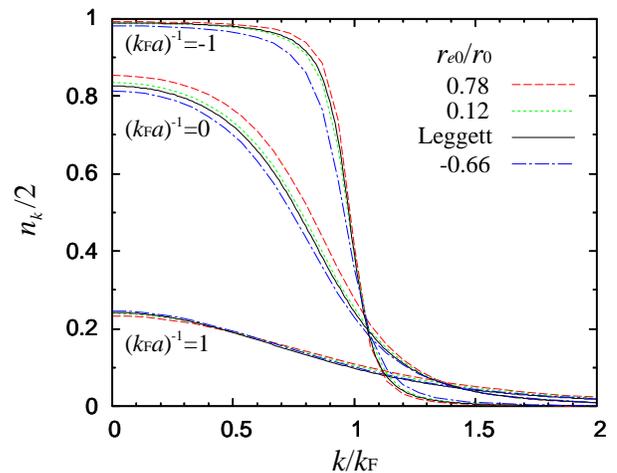}\end{center}
\caption{(Color online) The momentum distribution $n_k/2$ of fermion atoms
for $(k_{\mathrm{F}}a)^{-1}=-1,0$, and 1 is presented for several values of
of the effective range $r_{e0}$ in the unitarity limit.
The density is set to be $k_{\rm F}r_0\simeq 0.31$.
\label{cap:nkhalf} }
\end{figure}

The momentum distribution $n_k/2=|v_k|^2/\cal{V}$ 
of fermion atoms (for spin-up or spin-down)
is shown in Fig. \ref{cap:nkhalf}, where $n_k$ being $2|v_k|^2/\cal{V}$.
Here we plot $n_k/2$ for $(k_{\rm F}a)^{-1}=-1,0$, and 1  
for the same parameter sets as in Fig. \ref{cap:gap3sqwb}.
In the near to far BEC limits of $(k_{\rm F}a)\agt 1$, 
all curves for different $r_{e0}$
almost coincide with the one for the contact potential.
However, from the unitarity regime $(k_{\rm F}a)^{-1}\sim0$ 
to the near BCS limit $(k_{\rm F}a)^{-1}\sim -1$, we observe 
significant effective range corrections.
The results for $(k_{\rm F}a)^{-1}=-1$ and 0 in this figure show that
corrections due to the negative effective range suppress $n_{k=0}$
and broaden the momentum distribution.
As discussed before, a large, negative $r_{e0}$, 
which corresponds to a weak atom-molecule coupling, favors the BCS state
on the BCS side. Therefore the Fermi surface is smeared and 
$n_k$ is broadened in this case.

\subsubsection{Condensate fraction}

We consider the influence of the effective range on the condensate
fraction in the BCS-BEC crossover. In a Fermi gas, the appearance of
superfluid order is related to the appearance of off-diagonal long
range order as found in the two-body density matrix. Following
\cite{Campbell1997,Ortiz2005,Salasnich2005},
we calculate the number of the condensed fermion pairs as
$N_0=\sum_{\vec{k}} |u_{\vec{k}}v_{\vec{k}}|^2$
and examine the influence of the nonzero effective range.

\begin{figure}
\begin{center}\includegraphics[%
  width=8cm,
  keepaspectratio]{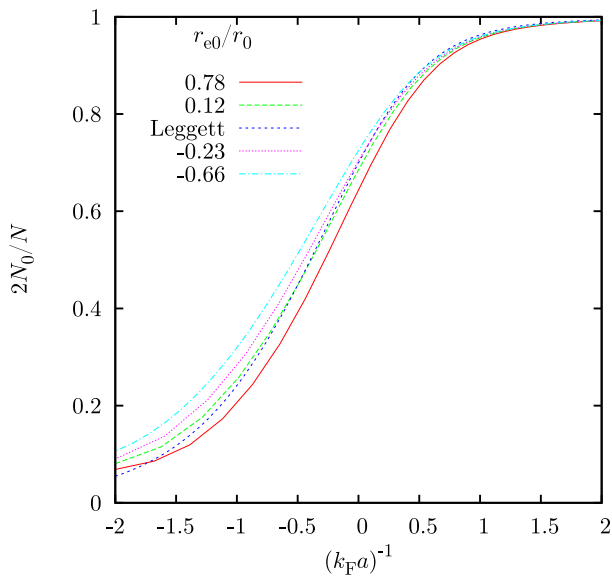}\end{center}
\caption{(Color online) The normalized zero temperature condensate fraction
$N_{0}/(N/2)$ as a function of the inverse coupling constant 
$(k_{\mathrm{F}}a)^{-1}$ 
for several values of the effective range $r_{e0}/r_{0}$.
The density is set to be $k_{\rm F}r_0\simeq 0.31$.
\label{cap:cfsqwb} }
\end{figure}

Figure \ref{cap:cfsqwb}
contains a plot of the normalized condensate fraction $N_0/(N/2)=2N_{0}/N$
of a Fermi gas 
as a function of the inverse coupling for the same parameter
sets as those used in Figs. \ref{cap:mu} and \ref{cap:gap-at-zero}.
Within the mean-field approximation, the normalized
molecular condensate fraction converges to unity in the BEC limit independent
of the detailed nature of the potential. On the other hand, when the
binding energy is decreased, the dimers dissociate more easily
and the condensate fraction correspondingly decreases and eventually vanishes
in the BCS limit. 
For a large and negative value of $r_{e0}$, the dissociation is suppressed
due to the weak atom-molecule coupling,
and thus the condensate fraction is enhanced as shown in this figure.

\section{Summary and conclusion}

We have presented a detailed study of the two-body bound state and
scattering properties of a simple solvable model potential with a barrier 
to model the narrow resonance phenomena in the BCS-BEC crossover
in atomic Fermi gases. We paid particular attention to the effect
of the nonzero effective range and its sign as the inverse scattering
length was tuned from large negative values to large positive ones.
The advantage of our simple single-channel potential 
model is that it explicitly displays
the relation between the potential resonance phenomena and the appearance
of negative effective range, which is predicted by two-channel models.

First, we studied the two-particle scattering problem with 
the square-well, square-barrier potential, and have clearly demonstrated 
the physical meaning of the negative effective range.
The effective range at resonance was carefully analyzed
and the analytical expression for the weak barrier case was also presented.
This expression shows that adding a barrier to a square-well potential
yields smaller values of the effective range.

We then applied this potential to the BCS-BEC crossover phenomena
within an effective single-channel potential model.
By solving the crossover equations self-consistently,
we studied the effects of both the positive and negative effective range
within a unified framework. We have observed that corrections
due to the negative effective range, in general,
appear as a maximum of the gap parameter at finite momentum
and a reduction of the chemical potential close to the resonance
and in the BCS region.
In view of the possibility of performing experiments 
on systems with narrow resonances,
we have discussed consequences of the nonzero effective range
on various measurable quantities 
\footnote[5]{The effective range corrections also affect, e.g., the equation
of state (EOS). The enhancement of the attractive interaction in the BCS side 
due to the negative effective range, which has been discussed in
Section \ref{sec:gap-at-zero}, would soften the EOS.
We have observed this softening within the numerical accuracy.
In this calculation, we assume a polytropic EOS,
$P\propto n^{\gamma+1}$, where $P$ is the pressure and $\gamma$ is the 
polytropic index, which is given by 
$\gamma=1+n(\partial^2\mu/\partial n^2)/(\partial\mu/\partial n)$.
However, the second derivative $(\partial^2\mu/\partial n^2)$ 
is hard to calculate accurately and our results of $\gamma$ are not
accurate enough to discuss quantitatively.}.
The results presented in this work will be helpful when one explores 
the BCS-BEC crossover for narrow resonances.

\begin{acknowledgments}
It is a pleasure to acknowledge C. J. Pethick for stimulating discussion 
and his support in completing this work.
We gratefully acknowledge C. E. Campbell for kindly 
sending us a copy of Ref. \cite{Campbell1997}.
We are also grateful to G. Baym for helpful discussion 
in the initial stage of this work.
One of the authors G. W. was partially supported by 
the Nishina Memorial Foundation and by the JSPS
Postdoctoral Program for Research Abroad.
\end{acknowledgments}

\bibliographystyle{apsrev}
\bibliography{./sqwb}

\begin{thebibliography}{36}
\expandafter\ifx\csname natexlab\endcsname\relax\def\natexlab#1{#1}\fi
\expandafter\ifx\csname bibnamefont\endcsname\relax
  \def\bibnamefont#1{#1}\fi
\expandafter\ifx\csname bibfnamefont\endcsname\relax
  \def\bibfnamefont#1{#1}\fi
\expandafter\ifx\csname citenamefont\endcsname\relax
  \def\citenamefont#1{#1}\fi
\expandafter\ifx\csname url\endcsname\relax
  \def\url#1{\texttt{#1}}\fi
\expandafter\ifx\csname urlprefix\endcsname\relax\def\urlprefix{URL }\fi
\providecommand{\bibinfo}[2]{#2}
\providecommand{\eprint}[2][]{\url{#2}}

\bibitem[{\citenamefont{Regal et~al.}(2004)\citenamefont{Regal, Greiner, and
  Jin}}]{Regal2004}
\bibinfo{author}{\bibfnamefont{C.~A.} \bibnamefont{Regal}},
  \bibinfo{author}{\bibfnamefont{M.}~\bibnamefont{Greiner}}, \bibnamefont{and}
  \bibinfo{author}{\bibfnamefont{D.~S.} \bibnamefont{Jin}},
  \bibinfo{journal}{Phys. Rev. Lett.} \textbf{\bibinfo{volume}{92}},
  \bibinfo{pages}{040403} (\bibinfo{year}{2004}).

\bibitem[{\citenamefont{Zwierlein et~al.}(2004)\citenamefont{Zwierlein, Stan,
  Schunck, Raupach, Kerman, and Ketterle}}]{Zwierlein2004a}
\bibinfo{author}{\bibfnamefont{M.~W.} \bibnamefont{Zwierlein}},
  \bibinfo{author}{\bibfnamefont{C.~A.} \bibnamefont{Stan}},
  \bibinfo{author}{\bibfnamefont{C.~H.} \bibnamefont{Schunck}},
  \bibinfo{author}{\bibfnamefont{S.~M.~F.} \bibnamefont{Raupach}},
  \bibinfo{author}{\bibfnamefont{A.~J.} \bibnamefont{Kerman}},
  \bibnamefont{and} \bibinfo{author}{\bibfnamefont{W.}~\bibnamefont{Ketterle}},
  \bibinfo{journal}{Phys. Rev. Lett.} \textbf{\bibinfo{volume}{92}},
  \bibinfo{pages}{120403} (\bibinfo{year}{2004}).

\bibitem[{\citenamefont{Kinast et~al.}(2004)\citenamefont{Kinast, Hemmer, Gehm,
  Turlapov, and Thomas}}]{Kinast2004}
\bibinfo{author}{\bibfnamefont{J.}~\bibnamefont{Kinast}},
  \bibinfo{author}{\bibfnamefont{S.~L.} \bibnamefont{Hemmer}},
  \bibinfo{author}{\bibfnamefont{M.~E.} \bibnamefont{Gehm}},
  \bibinfo{author}{\bibfnamefont{A.}~\bibnamefont{Turlapov}}, \bibnamefont{and}
  \bibinfo{author}{\bibfnamefont{J.~E.} \bibnamefont{Thomas}},
  \bibinfo{journal}{Phys. Rev. Lett.} \textbf{\bibinfo{volume}{92}},
  \bibinfo{pages}{150402} (\bibinfo{year}{2004}).

\bibitem[{\citenamefont{Bartenstein
  et~al.}(2004{\natexlab{a}})\citenamefont{Bartenstein, Altmeyer, Riedl,
  Jochim, Chin, Denschlag, and Grimm}}]{Bartenstein2004col}
\bibinfo{author}{\bibfnamefont{M.}~\bibnamefont{Bartenstein}},
  \bibinfo{author}{\bibfnamefont{A.}~\bibnamefont{Altmeyer}},
  \bibinfo{author}{\bibfnamefont{S.}~\bibnamefont{Riedl}},
  \bibinfo{author}{\bibfnamefont{S.}~\bibnamefont{Jochim}},
  \bibinfo{author}{\bibfnamefont{C.}~\bibnamefont{Chin}},
  \bibinfo{author}{\bibfnamefont{J.~H.} \bibnamefont{Denschlag}},
  \bibnamefont{and} \bibinfo{author}{\bibfnamefont{R.}~\bibnamefont{Grimm}},
  \bibinfo{journal}{Phys. Rev. Lett.} \textbf{\bibinfo{volume}{92}},
  \bibinfo{pages}{203201} (\bibinfo{year}{2004}{\natexlab{a}}).

\bibitem[{\citenamefont{Chin et~al.}(2004)\citenamefont{Chin, Bartenstein,
  Altmeyer, Riedl, Jochim, Denschlag, and Grimm}}]{Chin2004}
\bibinfo{author}{\bibfnamefont{C.}~\bibnamefont{Chin}},
  \bibinfo{author}{\bibfnamefont{M.}~\bibnamefont{Bartenstein}},
  \bibinfo{author}{\bibfnamefont{A.}~\bibnamefont{Altmeyer}},
  \bibinfo{author}{\bibfnamefont{S.}~\bibnamefont{Riedl}},
  \bibinfo{author}{\bibfnamefont{S.}~\bibnamefont{Jochim}},
  \bibinfo{author}{\bibfnamefont{J.~H.} \bibnamefont{Denschlag}},
  \bibnamefont{and} \bibinfo{author}{\bibfnamefont{R.}~\bibnamefont{Grimm}},
  \bibinfo{journal}{Science} \textbf{\bibinfo{volume}{305}},
  \bibinfo{pages}{1128} (\bibinfo{year}{2004}).

\bibitem[{\citenamefont{Zwierlein et~al.}(2005)\citenamefont{Zwierlein,
  Abo-Shaeer, Schirotzek, Schunck, and Ketterle}}]{Zwierlein2005b}
\bibinfo{author}{\bibfnamefont{M.~W.} \bibnamefont{Zwierlein}},
  \bibinfo{author}{\bibfnamefont{J.~R.} \bibnamefont{Abo-Shaeer}},
  \bibinfo{author}{\bibfnamefont{A.}~\bibnamefont{Schirotzek}},
  \bibinfo{author}{\bibfnamefont{C.~H.} \bibnamefont{Schunck}},
  \bibnamefont{and} \bibinfo{author}{\bibfnamefont{W.}~\bibnamefont{Ketterle}},
  \bibinfo{journal}{Nature} \textbf{\bibinfo{volume}{435}},
  \bibinfo{pages}{1047} (\bibinfo{year}{2005}).

\bibitem[{\citenamefont{Eagles}(1969)}]{Eagles1969b}
\bibinfo{author}{\bibfnamefont{D.~M.} \bibnamefont{Eagles}},
  \bibinfo{journal}{Phys. Rev.} \textbf{\bibinfo{volume}{186}},
  \bibinfo{pages}{456} (\bibinfo{year}{1969}).

\bibitem[{\citenamefont{Leggett}(1980{\natexlab{a}})}]{Leggett1980}
\bibinfo{author}{\bibfnamefont{A.~J.} \bibnamefont{Leggett}},
  \emph{\bibinfo{title}{{\rm in} {M}odern {T}rends in the {T}heory of
  {C}ondensed {M}atter}} (\bibinfo{publisher}{Springer Verlag},
  \bibinfo{year}{1980}{\natexlab{a}}), p.~\bibinfo{pages}{13}.

\bibitem[{\citenamefont{Leggett}(1980{\natexlab{b}})}]{Leggett1980b}
\bibinfo{author}{\bibfnamefont{A.~J.} \bibnamefont{Leggett}},
  \bibinfo{journal}{J. de Phys., Coll.} \textbf{\bibinfo{volume}{7}},
  \bibinfo{pages}{C7} (\bibinfo{year}{1980}{\natexlab{b}}).

\bibitem[{\citenamefont{Bruun and Pethick}(2004)}]{Bruun2004a}
\bibinfo{author}{\bibfnamefont{G.~M.} \bibnamefont{Bruun}} \bibnamefont{and}
  \bibinfo{author}{\bibfnamefont{C.~J.} \bibnamefont{Pethick}},
  \bibinfo{journal}{Phys. Rev. Lett.} \textbf{\bibinfo{volume}{92}},
  \bibinfo{pages}{140404} (\bibinfo{year}{2004}).

\bibitem[{\citenamefont{Petrov et~al.}(2005)\citenamefont{Petrov, Salomon, and
  Shlyapnikov}}]{Petrov2004b}
\bibinfo{author}{\bibfnamefont{D.~S.} \bibnamefont{Petrov}},
  \bibinfo{author}{\bibfnamefont{C.}~\bibnamefont{Salomon}}, \bibnamefont{and}
  \bibinfo{author}{\bibfnamefont{G.~V.} \bibnamefont{Shlyapnikov}},
  \bibinfo{journal}{Phys. Rev. A} \textbf{\bibinfo{volume}{71}},
  \bibinfo{pages}{012708} (\bibinfo{year}{2005}).

\bibitem[{\citenamefont{Bartenstein
  et~al.}(2004{\natexlab{b}})\citenamefont{Bartenstein, Altmeyer, Riedl,
  Jochim, Chin, Denschlag, and Grimm}}]{Bartenstein2004a}
\bibinfo{author}{\bibfnamefont{M.}~\bibnamefont{Bartenstein}},
  \bibinfo{author}{\bibfnamefont{A.}~\bibnamefont{Altmeyer}},
  \bibinfo{author}{\bibfnamefont{S.}~\bibnamefont{Riedl}},
  \bibinfo{author}{\bibfnamefont{S.}~\bibnamefont{Jochim}},
  \bibinfo{author}{\bibfnamefont{C.}~\bibnamefont{Chin}},
  \bibinfo{author}{\bibfnamefont{J.~H.} \bibnamefont{Denschlag}},
  \bibnamefont{and} \bibinfo{author}{\bibfnamefont{R.}~\bibnamefont{Grimm}},
  \bibinfo{journal}{Phys. Rev. Lett.} \textbf{\bibinfo{volume}{92}},
  \bibinfo{pages}{120401} (\bibinfo{year}{2004}{\natexlab{b}}).

\bibitem[{\citenamefont{Perali et~al.}(2004)\citenamefont{Perali, Pieri,
  Pisani, and Strinati}}]{Perali2004a}
\bibinfo{author}{\bibfnamefont{A.}~\bibnamefont{Perali}},
  \bibinfo{author}{\bibfnamefont{P.}~\bibnamefont{Pieri}},
  \bibinfo{author}{\bibfnamefont{L.}~\bibnamefont{Pisani}}, \bibnamefont{and}
  \bibinfo{author}{\bibfnamefont{G.~C.} \bibnamefont{Strinati}},
  \bibinfo{journal}{Phys. Rev. Lett} \textbf{\bibinfo{volume}{92}},
  \bibinfo{pages}{220404} (\bibinfo{year}{2004}).

\bibitem[{\citenamefont{Simonucci et~al.}(2005)\citenamefont{Simonucci, Pieri,
  and Strinati}}]{Simonucci2005}
\bibinfo{author}{\bibfnamefont{S.}~\bibnamefont{Simonucci}},
  \bibinfo{author}{\bibfnamefont{P.}~\bibnamefont{Pieri}}, \bibnamefont{and}
  \bibinfo{author}{\bibfnamefont{G.~C.} \bibnamefont{Strinati}},
  \bibinfo{journal}{Europhys. Lett.} \textbf{\bibinfo{volume}{69}},
  \bibinfo{pages}{713} (\bibinfo{year}{2005}).

\bibitem[{\citenamefont{Heiselberg}(2004)}]{Heiselberg2004}
\bibinfo{author}{\bibfnamefont{H.}~\bibnamefont{Heiselberg}},
  \bibinfo{journal}{Phys. Rev. Lett.} \textbf{\bibinfo{volume}{93}},
  \bibinfo{pages}{040402} (\bibinfo{year}{2004}).

\bibitem[{\citenamefont{Hu et~al.}(2004)\citenamefont{Hu, Minguzzi, Liu, and
  Tosi}}]{Hu2004}
\bibinfo{author}{\bibfnamefont{H.}~\bibnamefont{Hu}},
  \bibinfo{author}{\bibfnamefont{A.}~\bibnamefont{Minguzzi}},
  \bibinfo{author}{\bibfnamefont{X.-J.} \bibnamefont{Liu}}, \bibnamefont{and}
  \bibinfo{author}{\bibfnamefont{M.~P.} \bibnamefont{Tosi}},
  \bibinfo{journal}{Phys. Rev. Lett.} \textbf{\bibinfo{volume}{93}},
  \bibinfo{pages}{190403} (\bibinfo{year}{2004}).

\bibitem[{\citenamefont{Andrenacci et~al.}(1999)\citenamefont{Andrenacci,
  Perali, Pieri, and Strinati}}]{Andrenacci1999}
\bibinfo{author}{\bibfnamefont{N.}~\bibnamefont{Andrenacci}},
  \bibinfo{author}{\bibfnamefont{A.}~\bibnamefont{Perali}},
  \bibinfo{author}{\bibfnamefont{P.}~\bibnamefont{Pieri}}, \bibnamefont{and}
  \bibinfo{author}{\bibfnamefont{G.~C.} \bibnamefont{Strinati}},
  \bibinfo{journal}{Phys. Rev. B} \textbf{\bibinfo{volume}{60}},
  \bibinfo{pages}{12410} (\bibinfo{year}{1999}).

\bibitem[{\citenamefont{Kokkelmans et~al.}(2002)\citenamefont{Kokkelmans,
  Milstein, Chiofalo, Walser, and Holland}}]{Kokkelmans2002}
\bibinfo{author}{\bibfnamefont{S.~J.~J.~M.~F.} \bibnamefont{Kokkelmans}},
  \bibinfo{author}{\bibfnamefont{J.~N.} \bibnamefont{Milstein}},
  \bibinfo{author}{\bibfnamefont{M.~L.} \bibnamefont{Chiofalo}},
  \bibinfo{author}{\bibfnamefont{R.}~\bibnamefont{Walser}}, \bibnamefont{and}
  \bibinfo{author}{\bibfnamefont{M.~J.} \bibnamefont{Holland}},
  \bibinfo{journal}{Phys. Rev. A} \textbf{\bibinfo{volume}{65}},
  \bibinfo{pages}{053617} (\bibinfo{year}{2002}).

\bibitem[{\citenamefont{Carlson et~al.}(2003)\citenamefont{Carlson, Chang,
  Pandharipande, and Schmidt}}]{Carlson2003}
\bibinfo{author}{\bibfnamefont{J.}~\bibnamefont{Carlson}},
  \bibinfo{author}{\bibfnamefont{S.~Y.} \bibnamefont{Chang}},
  \bibinfo{author}{\bibfnamefont{V.~R.} \bibnamefont{Pandharipande}},
  \bibnamefont{and} \bibinfo{author}{\bibfnamefont{K.~E.}
  \bibnamefont{Schmidt}}, \bibinfo{journal}{Phys. Rev. Lett.}
  \textbf{\bibinfo{volume}{91}}, \bibinfo{pages}{050401}
  (\bibinfo{year}{2003}).

\bibitem[{\citenamefont{Parish et~al.}(2005)\citenamefont{Parish, Mihaila,
  Timmermans, Blagoev, and Littlewood}}]{Parish2005a}
\bibinfo{author}{\bibfnamefont{M.~M.} \bibnamefont{Parish}},
  \bibinfo{author}{\bibfnamefont{B.}~\bibnamefont{Mihaila}},
  \bibinfo{author}{\bibfnamefont{E.~M.} \bibnamefont{Timmermans}},
  \bibinfo{author}{\bibfnamefont{K.~B.} \bibnamefont{Blagoev}},
  \bibnamefont{and} \bibinfo{author}{\bibfnamefont{P.~B.}
  \bibnamefont{Littlewood}}, \bibinfo{journal}{Phys. Rev. B}
  \textbf{\bibinfo{volume}{71}}, \bibinfo{pages}{064513}
  (\bibinfo{year}{2005}).

\bibitem[{\citenamefont{{De Palo} et~al.}(2004)\citenamefont{{De Palo},
  Chiofalo, Holland, and Kokkelmans}}]{Palo2004}
\bibinfo{author}{\bibfnamefont{S.}~\bibnamefont{{De Palo}}},
  \bibinfo{author}{\bibfnamefont{M.~L.} \bibnamefont{Chiofalo}},
  \bibinfo{author}{\bibfnamefont{M.~J.} \bibnamefont{Holland}},
  \bibnamefont{and} \bibinfo{author}{\bibfnamefont{S.~J. J. M.~F.}
  \bibnamefont{Kokkelmans}}, \bibinfo{journal}{Phys. Lett. A}
  \textbf{\bibinfo{volume}{327}}, \bibinfo{pages}{490} (\bibinfo{year}{2004}).

\bibitem[{\citenamefont{Diener and Ho}(2004{\natexlab{a}})}]{Diener2004b}
\bibinfo{author}{\bibfnamefont{R.}~\bibnamefont{Diener}} \bibnamefont{and}
  \bibinfo{author}{\bibfnamefont{T.-L.} \bibnamefont{Ho}},
  \bibinfo{journal}{cond-mat/0405174}  (\bibinfo{year}{2004}{\natexlab{a}}).

\bibitem[{\citenamefont{Holland et~al.}(2001)\citenamefont{Holland, Kokkelmans,
  Chiofalo, and R.}}]{Holland2001}
\bibinfo{author}{\bibfnamefont{M.}~\bibnamefont{Holland}},
  \bibinfo{author}{\bibfnamefont{S.~J. J. M.~F.} \bibnamefont{Kokkelmans}},
  \bibinfo{author}{\bibfnamefont{M.~L.} \bibnamefont{Chiofalo}},
  \bibnamefont{and} \bibinfo{author}{\bibnamefont{R.}}, \bibinfo{journal}{Phys.
  Rev. Lett.} \textbf{\bibinfo{volume}{87}}, \bibinfo{pages}{120406}
  (\bibinfo{year}{2001}).

\bibitem[{\citenamefont{Blatt and Jackson}(1949)}]{Blatt1949}
\bibinfo{author}{\bibfnamefont{J.~M.} \bibnamefont{Blatt}} \bibnamefont{and}
  \bibinfo{author}{\bibfnamefont{J.~D.} \bibnamefont{Jackson}},
  \bibinfo{journal}{Phys. Rev.} \textbf{\bibinfo{volume}{76}},
  \bibinfo{pages}{18} (\bibinfo{year}{1949}).

\bibitem[{\citenamefont{Bethe}(1949)}]{Bethe1949}
\bibinfo{author}{\bibfnamefont{H.~A.} \bibnamefont{Bethe}},
  \bibinfo{journal}{Phys. Rev.} \textbf{\bibinfo{volume}{76}},
  \bibinfo{pages}{38} (\bibinfo{year}{1949}).

\bibitem[{\citenamefont{Fl\"ugge}(1999)}]{Flugge1999}
\bibinfo{author}{\bibfnamefont{S.}~\bibnamefont{Fl\"ugge}},
  \emph{\bibinfo{title}{Practical {Q}uantum {M}echanics}}
  (\bibinfo{publisher}{{S}pringer}, \bibinfo{year}{1999}).

\bibitem[{\citenamefont{Diener and Ho}(2004{\natexlab{b}})}]{Diener2004a}
\bibinfo{author}{\bibfnamefont{R.~B.} \bibnamefont{Diener}} \bibnamefont{and}
  \bibinfo{author}{\bibfnamefont{T.-L.} \bibnamefont{Ho}},
  \bibinfo{journal}{cond-mat/0404517}  (\bibinfo{year}{2004}{\natexlab{b}}).

\bibitem[{\citenamefont{Haussmann}(1993)}]{Haussmann1993}
\bibinfo{author}{\bibfnamefont{R.}~\bibnamefont{Haussmann}},
  \bibinfo{journal}{Z. Phys. B} \textbf{\bibinfo{volume}{91}},
  \bibinfo{pages}{291} (\bibinfo{year}{1993}).

\bibitem[{\citenamefont{Marini et~al.}(1998)\citenamefont{Marini, Pistolesi,
  and Strinati}}]{Marini1998}
\bibinfo{author}{\bibfnamefont{M.}~\bibnamefont{Marini}},
  \bibinfo{author}{\bibfnamefont{F.}~\bibnamefont{Pistolesi}},
  \bibnamefont{and} \bibinfo{author}{\bibfnamefont{G.~C.}
  \bibnamefont{Strinati}}, \bibinfo{journal}{Eur. Phys. J. B}
  \textbf{\bibinfo{volume}{1}}, \bibinfo{pages}{151} (\bibinfo{year}{1998}).

\bibitem[{\citenamefont{Papenbrock and Bertsch}(1999)}]{Papenbrock1999}
\bibinfo{author}{\bibfnamefont{T.}~\bibnamefont{Papenbrock}} \bibnamefont{and}
  \bibinfo{author}{\bibfnamefont{G.~F.} \bibnamefont{Bertsch}},
  \bibinfo{journal}{Phys. Rev. C} \textbf{\bibinfo{volume}{59}},
  \bibinfo{pages}{2052} (\bibinfo{year}{1999}).

\bibitem[{not({\natexlab{a}})}]{note-1}
\bibinfo{note}{Under the assumption of isotropic gap, $\Delta_{\vec
  k}=\Delta(k)$, we can show from the gap equation that the minimum of $E_k$
  shifts from $k=0$ when the chemical potential equals to
  $(m\Delta(0)/\hbar^2)\partial^2_k\Delta(0)$ and the maximum of $\psi(k)$
  shifts from $k=0$ when $\mu=0$.}

\bibitem[{not({\natexlab{b}})}]{note-2}
\bibinfo{note}{In the BCS limit, the pair wave function $\psi(k)=\Delta_{\vec
  k}/(2E_{\vec k})$ is very narrow around $k=k_{\rm F}$ and the gap equation is
  hard to solve accurately by our code. This is why we do not show the result
  for $r_{e0}/r_0=-1.20$ at $(k_{\rm F}a)^{-1}\alt -1.1$ in this figure and
  Fig. \ref{cap:gap-at-zero}.}

\bibitem[{\citenamefont{Bruun}(2004)}]{Bruun2004b}
\bibinfo{author}{\bibfnamefont{G.~M.} \bibnamefont{Bruun}},
  \bibinfo{journal}{Phys. Rev. A} \textbf{\bibinfo{volume}{70}},
  \bibinfo{pages}{053602} (\bibinfo{year}{2004}).

\bibitem[{\citenamefont{Campbell}(1997)}]{Campbell1997}
\bibinfo{author}{\bibfnamefont{C.~E.} \bibnamefont{Campbell}},
  \emph{\bibinfo{title}{{\rm in} {C}ondensed {M}atter {T}heories}}
  (\bibinfo{publisher}{Nova Science}, \bibinfo{year}{1997}),
  vol.~\bibinfo{volume}{12}, p. \bibinfo{pages}{131}.

\bibitem[{\citenamefont{Ortiz and Dukelsky}(2005)}]{Ortiz2005}
\bibinfo{author}{\bibfnamefont{G.}~\bibnamefont{Ortiz}} \bibnamefont{and}
  \bibinfo{author}{\bibfnamefont{J.}~\bibnamefont{Dukelsky}},
  \bibinfo{journal}{Phys. Rev. A} \textbf{\bibinfo{volume}{72}},
  \bibinfo{pages}{043611} (\bibinfo{year}{2005}).

\bibitem[{\citenamefont{Salasnich et~al.}(2005)\citenamefont{Salasnich, Manini,
  and Parola}}]{Salasnich2005}
\bibinfo{author}{\bibfnamefont{L.}~\bibnamefont{Salasnich}},
  \bibinfo{author}{\bibfnamefont{N.}~\bibnamefont{Manini}}, \bibnamefont{and}
  \bibinfo{author}{\bibfnamefont{A.}~\bibnamefont{Parola}},
  \bibinfo{journal}{Phys. Rev. A} \textbf{\bibinfo{volume}{72}},
  \bibinfo{pages}{023621} (\bibinfo{year}{2005}).

\end{thebibliography}

\end{document}